# Secure Lossy Source Coding for Some Classes of Helper and Gray-Wyner Models

Meryem Benammar Abdellatif Zaidi

**Abstract**

In this work, we investigate two source coding models, a *Helper* problem and a *Gray-Wyner* problem, under equivocation constraints. Specifically, in the Helper problem, an encoder communicates with a legitimate receiver through a noise-free rate-limited public link as well as a noise-free rate-limited private link; and an external passive eavesdropper intercepts every information that is sent on the public link. We study two classes of this model: i) when a pair of arbitrarily correlated discrete memoryless sources is to be encoded such that one component has to be recovered lossily at the legitimate receiver while the equivocation about both components at the eavesdropper must be maintained no smaller than some prescribed level; and ii) when the legitimate receiver reproduces both components, one of which, that is recovered losslessly, has to be concealed from the eavesdropper to some equivocation level. For both classes of Helper problems, we establish single-letter characterizations of optimal rate-distortion-equivocation tradeoffs in the discrete memoryless case. Next, we extend our results to the case of two legitimate receivers, i.e., Gray-Wyner type network model with equivocation constraints. Here, two legitimate receivers are connected to the encoder each through a dedicated error-free private link as well as a common error-free public link; and an external passive eavesdropper overhears on the public link. We study two classes of this model that are extensions of the aforementioned instances of Helper problems to the case of two receivers. For each of the two classes, we establish a single-letter characterization of the optimal rate-distortion-equivocation region. Throughout the paper, the analysis sheds light on the role of the private links, and we illustrate the results by computing them for some binary examples. Also, we make some meaningful connections, e.g., with problems of secret-sharing and encryption.

## I. Introduction

Since its introduction by Shannon in the fifties, information theoretic security has motivated extensive research, in diverse directions. However, much of this research is directed towards *channel coding* models, e.g., the wiretap channel [1], the broadcast channel with confidential messages [2]

M. Benammar is with the Mathematical and Algorithmic Sciences Lab., Huawei Technologies France, Boulogne-Billancourt, 92100, France. A. Zaidi was with Université Paris-Est, Champs-sur-Marne, 77454, France, and is on leave at the Mathematical and Algorithmic Sciences Lab., Huawei Technologies France, Boulogne-Billancourt, 92100. Emails: meryem.benammar@huawei.com, abdellatif.zaidi@huawei.com, abdellatif.zaidi@u-pem.fr.

and numerous extensions of these works to vector wiretap channels [3]–[6] as well as various multi-terminal settings such as the multiple access wiretap channel [7]–[10], the relay channel [11], the interference channel [12] and X networks [13], [14] (the reader may refer to [15] for a review of many other related contributions). Secrecy-oriented *source coding* models, however, have attracted less interest comparatively, and are way less well understood in general. Perhaps, this is due to the source-channel coding separation which is often applied, though generally suboptimal; or the folklore interpretation that information secrecy is generally facilitated by injecting additional "noise" on the channels to eavesdroppers.

One important source coding model with secrecy constraints, which was studied by Yamamoto in [16], is shown in Figure 1. In this model, an encoder observes a pair of arbitrarily correlated discrete-memoryless (DM) sources $(S_1^n, S_2^n)$, and communicates with a receiver over a noise-free rate-limited public communication channel of capacity $R$. An external eavesdropper intercepts all information on the public link. The goal of the communication is to allow the legitimate receiver to reproduce the source component $S_1^n$ lossily, i.e., to within some desired average fidelity level $D$, while maintaining the equivocation at the eavesdropper about the source component $S_2^n$ no smaller than some prescribed level $\Delta$. Yamamoto characterized in [16] the region of optimal tradeoff among compression ratio $R$, average distortion $D$ and equivocation $\Delta$. From a practical viewpoint, the model of Figure 1 can be useful to model scenarios in which some critical information (e.g., medical information about a patient) or specific feature of a source needs to be kept secret from adversarial interceptions.

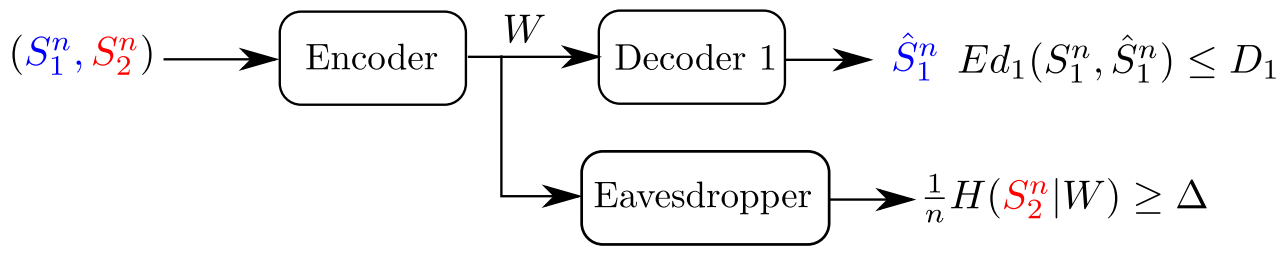

Fig. 1: Yamamoto's model for secure lossy source coding

In Yamamoto's setting, since the entire information that is transmitted to the legitimate receiver gets intercepted by the eavesdropper, perfect secrecy sometimes comes at the price of no transmission at all (e.g., think about the case in which $S_1^n = S_2^n = S^n$ and $S^n$ is to be recovered losslessly at the legitimate receiver). In this work, we investigate two related models in which legitimate receivers also have dedicated links on which they communicate with the encoder, and that cannot be intercepted by the eavesdropper. This is in accordance with the observation that secrecy is generally obtained by creating some *advantage* for legitimate receivers over eavesdroppers or adversaries. Specifically, we study a Helper problem and a Gray-Wyner model, both with equivocation constraints. The two models are described in the following two sections, together with our main contributions for each of them.

### *A. Helper Model and Contributions*

Consider the model shown in Figure 2. In this model, an encoder observes a pair of arbitrarily correlated memoryless sources $(S_0^n, S_1^n)$ and communicates with a legitimate receiver through a noise-free rate-limited public link as well as a noise-free rate-limited private link, in the presence of an external passive eavesdropper which intercepts every information that is sent on the public link. For convenience, and with some abuse in the terminology, we interpret the transmission on the private link as being enabled by a Helper node that observes the same sources and whose role is to facilitate the communication between the legitimate transmitter-receiver pair. (Formally, this is then an instance of the more general Helper problem in which the helper node observes a distinct, possibly correlated, source).

We study two classes of this model. In the first class, shown in Figure 2a, the communication should allow the legitimate receiver to reproduce only the source component $S_1^n$ lossily, to within some desired average distortion $D_1$, while maintaining the equivocation at the eavesdropper about both sources $(S_0^n, S_1^n)$ no smaller than some prescribed level $\Delta$. In this model, the encoder sends descriptions of the pair $(S_0^n, S_1^n)$ on both links. Let $R$ denote the rate of the compression index (or message) that the encoder sends on the public link, and $R_1$ that on the private link. Depending on the values of $D_1$ and $\Delta$, the communication strikes a good balance among revealing enough information to the legitimate receiver using both links and leaking only minimum information to the eavesdropper on the public link (about the sources themselves, not the compression indices).

In the second class, shown in Figure 2b, the legitimate receiver wants to reproduce both sources – the source component $S_0^n$ is reconstructed in a lossless manner and the source component $S_1^n$ is reconstructed in a lossy manner, to within some average distortion $D_1$. However, in this model, only the source $S_0^n$ represents sensitive information, and the communication is required to maintain the equivocation about it at the eavesdropper bounded from below by some prescribed level $\Delta$.

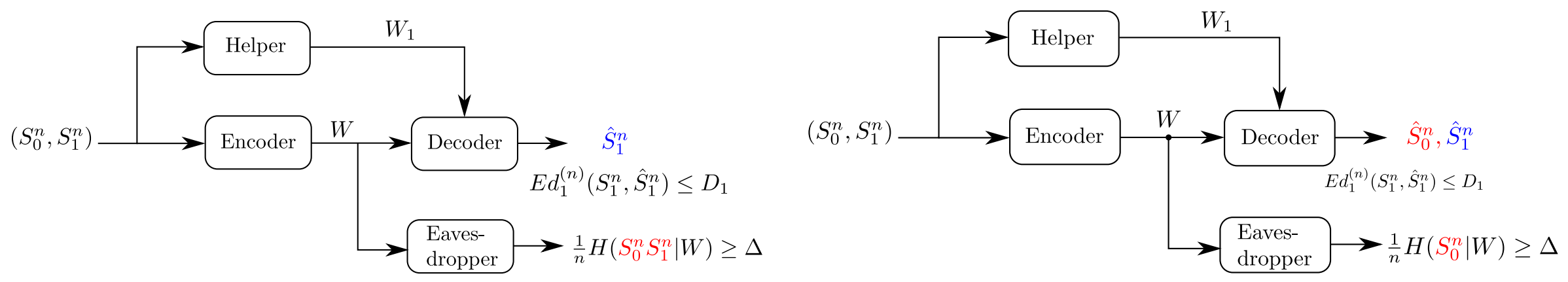

(a) Equivocation on both sources

(b) Reproducing both sources

Fig. 2: Two classes of Helper problems with equivocation constraints

Although seemingly similar, the aforementioned two classes of Helper problems with equivocation constraints exhibit different coding challenges. For example, while in the model of Figure 2a the source component $S_0^n$, which has to be kept secret to some level, needs *in principle* not be transmitted, it is required by the legitimate receiver in the model of Figure 2b. Conversely, there is a stronger restriction on the use of the public link in the model of Figure 2a, since, in the model of Figure 2b, descriptions of $S_1^n$ that leak only minimum information about $S_0^n$ can still be transmitted on the public link. However, in the investigation of both models, equally important is the understanding of the role of the private link. We will show that the two models require different ways of using this link optimally. In fact, by opposition to Yamamoto's setting of Figure 1, the presence of this link in the models that we study makes the derivation of optimal results (in terms of rate-distortion-equivocation) rather challenging.

For both classes of Helper problems, we establish single-letter characterizations of optimal rate-distortion-equivocation tradeoffs in the discrete memoryless case. For the model of Figure 2a, the optimal coding scheme is one in which the encoder sends a description of the pair $(S_0^n, S_1^n)$ on the public link, which is then utilized as side information at both encoder and decoder to send another description of $(S_0^n, S_1^n)$ on the private link. It is important to note that, in doing so, although the source component $S_0^n$ is not required by the legitimate receiver and is to be kept secret (to some level) at the eavesdropper, the description that is sent on the public link is arbitrarily correlated with $S_0^n$, an aspect that appears also in Yamamoto's optimal coding [16, Theorem 1] for the model of Figure 1 and is useful to enabling smaller equivocations at the eavesdropper.

For the model of Figure 2b, an optimal coding scheme is one in which the encoder describes the entire $S_0^n$ through the private link and uses the remaining of the capacity of this link, as well as the public link, to describe the source component $S_1^n$ to the legitimate receiver. In doing so, it takes into account $S_0^n$ as side information that is available at both encoder and decoder. Recalling that the source $S_1^n$ is arbitrarily correlated with the source $S_0^n$ which is to be kept secret from the eavesdropper, the analysis of the equivocation of this scheme requires showing that the compression index of $S_1^n$ that is sent on the public link leaks no information about $S_0^n$, which we prove through a here established new counting lemma. Furthermore, we establish some meaningful connections with the problem of secret-sharing and encryption, by showing that the private link can be utilized optimally in, essentially, two different ways, i) revealing the part of the sources that is to be kept secret to the legitimate receiver, or ii) enabling the encoder and legitimate receiver to share a common secret key which they then utilize to encrypt the part of the source that is to be secured on the public link, using a one time-pad approach which is reminiscent of encryption in conventional cryptosystems [17]. Finally, for both classes of models, we illustrate the results by computing the established regions for some binary examples.

### *B. Gray-Wyner Model and Contributions*

Next, we study generalizations of the aforementioned models to the case of two legitimate receivers, i.e., a Gray-Wyner network [18] with equivocation constraints, as shown in Figure 3. Here, the transmitter communicates with two legitimate receivers, each over a dedicated noise-free rate-limited private link as well as a common noise-free rate-limited public link; while an external passive eavesdropper overhears the public link. We study two classes of models. In the first class, shown in Figure 3a, the encoder observes a pair of arbitrarily correlated memoryless sources $(S_1^n, S_2^n)$ and wishes to describe the source $S_1^n$ to the first receiver and the source $S_2^n$ to the second receiver. Both reconstructions are performed in a lossy manner, to within some desired average distortions $D_1$ and $D_2$ respectively; and the communication should maintain the equivocation at the eavesdropper about both sources no smaller than some prescribed level $\Delta$. In this model, the imposed secrecy constraint may lead (depending on the value of $\Delta$) to rate redundancy on the private links; and the communication should strike a good balance among the compression rate $R_1$ on the private link to the first receiver, the compression rate $R_2$ on the private link to the second receiver, the compression rate $R$ on the public link, the desired distortion pair $(D_1, D_2)$ and the prescribed equivocation level $\Delta$.

In the second class of the models that we study, shown in Figure 3b, we introduce a third source component, $S_0^n$, that may represent the common information between $S_1^n$ and $S_2^n$ – the source triple $(S_0^n, S_1^n, S_2^n)$ are arbitrarily correlated, however, and $S_1^n$ and $S_2^n$ need not be independent conditionally on $S_0^n$. Also, we restrict the secrecy constraint to be only about $S_0^n$. Specifically, both receivers want to reproduce $S_0^n$ in a lossless manner; and the equivocation at the eavesdropper about this component should be no smaller than some desired level $\Delta$. In addition, Receiver 1 wants to reconstruct the source $S_1^n$ lossily to within average distortion $D_1$, and the second receiver wants to reconstruct the source $S_2^n$ lossily to within average distortion $D_2$. Recalling that for the original Gray-Wyner network without secrecy constraints [18] optimal coding necessitates that the source component $S_0^n$, and also any common information of the sources $S_1^n$ and $S_2^n$ conditionally on $S_0^n$, should be transmitted on the public link, it is clear that the imposed secrecy constraint changes drastically the problem. In fact, depending on the desired equivocation level, there is a tension among saving rate by sending $S_0^n$ on the public link and revealing only minimum information about it to the eavesdropper.

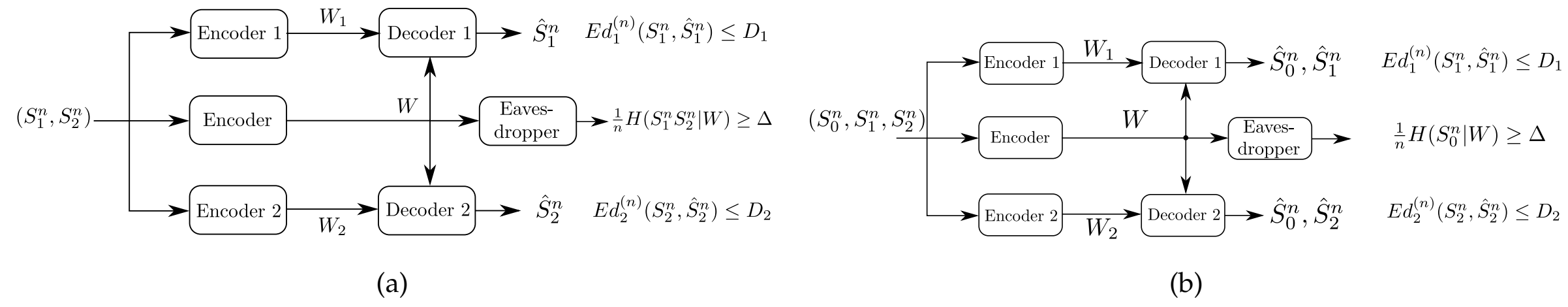

Fig. 3: Two classes of Gray-Wyner problems with equivocation constraints

For both classes of Gray-Wyner problems, we establish single-letter characterizations of optimal rate-distortion-equivocation tradeoffs in the discrete memoryless case. For the model of Figure 3b, perhaps non-intuitively, the analysis shows that the price of the imposed secrecy constraint is to send the source component $S_0^n$ on both private links, even though this entails some rate redundancy that otherwise could be saved if there were no eavesdropper.

The models of Figure 2 and of Figure 3 may be useful to model communication scenarios in which adversaries cannot intercept data on certain communication resources, e.g., time or frequency bands. For example, in the context of on-line caching, caching phases may be designed by the system operator on resources (time, frequency, code) that users cannot know apriori. Conversely, delivery phases usually happen during large traffic periods; and, so, transmission during this phase is more likely to be intercepted comparatively. Another driving problem, of interest in this setting, is a sensor network setting in which multiple sensors observe an underlying phenomenon that needs to be reconstructed at some access point. While some of the network connections may be secured (possibly wired), others might be wireless and so subjected to eavesdropping.

### C. Related Work

Most related to this contribution are works on secure transmission from a source coding perspective. A pioneering line of work in this aspect targets exploiting correlation among observations or sources at multiple terminals to generate secret common randomness which then may be utilized to encrypt data that is independent from the observations. The reader may refer to Maurer [19], Ahlswede-Csiszár [20] and Csiszár-Narayan [21] seminal papers for more on this line of work. In [22], Prabhakaran and Ramchandran study a setup in which the correlated observations or sources are themselves the data that needs to be communicated. Specifically, they study a Slepian-Wolf setting [23] in the presence of an eavesdropper with a correlated observation in which the goal is to convey the transmitter's source observation losslessly to the receiver while revealing the least amount of information about it to an eavesdropper that overhears the transmission. They characterize the optimal (minimum) leakage rate in this case. Perhaps of some importance is the observation, revealed therein through a simple example, that, unlike the setting without secrecy constraints, i.e., the standard Slepian-Wolf model [23], knowledge of decoder's side information at the transmitter is generally beneficial to enable higher equivocation at the eavesdropper. Also, Prabhakaran and Ramchandran consider a variation of the model in which the transmitter and receiver are allowed to interact over multiple rounds and they establish a lower bound on the leakage rate through connection with secret-sharing. In [22], no communication rate constraints are imposed on the sources. In [24] (see also [25]), Gunduz *et al.* study a related setting in which communication rate constraints are put on the sources. They consider both cases of coded and non-coded side information at the legitimate receiver. In the case of uncoded side information at the legitimate receiver, a complete characterization of the rate-equivocation region is established; and in the case of coded side information, bounds on this region, that do not match in general, are provided. In [26], Tandon *et al.* study a secure lossless source coding problem with a rate-limited Helper. In this work, they also consider a variation of the model in which the encoder has access to the coded output of the Helper. For both cases, they characterize the optimal rate-equivocation region. Other related works can be found in [27]–[30].

### D. Outline and Notation

An outline of the remainder of this paper is as follows. Section II is dedicated to the Helper models with equivocation constraints of Figure 2. After some formal definitions of the studied system models in Section II-A, we characterize the associated optimal rate-distortion equivocation regions in subsection II-B and subsection II-C respectively. Section III is devoted to the Gray-Wyner models with equivocation constraints of Figure 3. We define the two models formally in Section III-A, and then we characterize the associated rate-distortion-equivocation regions in Section III-B and Section III-C respectively. Sections II and III also contain some binary examples, as well as some useful discussions. The proofs are relegated to the appendices.

Throughout the paper we use the following notations. The term p.m.f. stands for probability mass function. Upper case letters are used to denote random variables, e.g., $X$; lower case letters are used to denote realizations of random variables, e.g., $x$; and calligraphic letters designate alphabets, i.e., $\mathcal{X}$. Vectors of length $n$ are denoted by $X^n = (X_1, \ldots, X_n)$, and $X_i^j$ is used to denote the sequence $(X_i, \ldots, X_j)$. The probability distribution of a random variable $X$ is denoted by $P_X(x) \triangleq \mathbb{P}(X = x)$. Sometimes, for convenience, we write it as $P_X$. We use the notation $\mathbb{E}_X[\cdot]$ to denote the expectation of random variable $X$. A probability distribution of a random variable $Y$ given $X$ is denoted by $P_{Y|X}$. The set of probability distributions defined on an alphabet $\mathcal{X}$ is denoted by $\mathcal{P}(\mathcal{X})$. The cardinality of a set $\mathcal{X}$ is denoted by $\|\mathcal{X}\|$. For random variables $X$, $Y$ and $Z$, the notation $X -\!\!\circ\!\!- Y -\!\!\circ\!\!- Z$ indicates that $X$, $Y$ and $Z$, in this order, form a Markov Chain, i.e., $P_{XYZ}(x, y, z) = P_Y(y)P_{X|Y}(x|y)P_{Z|Y}(z|y)$. The set $\mathcal{T}_{[X]}^{(n)}$ denotes the set of sequences strongly typical with respect to the probability distribution $P_X$ and the set $\mathcal{T}_{[X|y^n]}^{(n)}$ denotes the set of sequences $x^n$ jointly typical with $y^n$ with respect to the joint p.m.f. $P_{XY}$. Throughout this paper, we use $h_2(\alpha)$ to denote the entropy of a Bernoulli $(\alpha)$ source, i.e., $h_2(\alpha) = -\alpha \log(\alpha) - (1 - \alpha) \log(1 - \alpha)$. Also, the indicator function is denoted by $\mathbb{1}(\cdot)$. For real-valued scalars $a$ and $b$, with $a \le b$, the notation $[a, b]$ means the set of reals that are larger or equal than $a$ and smaller or equal $b$. For integers $i \le j$, $[i : j]$ denotes the set of integers comprised between $i$ and $j$, i.e., $[i : j] = \{i, i + 1, \ldots, j\}$. Finally, throughout the paper, logarithms are taken to base 2.

## II. Secure Lossy Helper Problems

In this section, we study the model of Figure 2. It is assumed that the alphabets $\mathcal{S}_0$, $\mathcal{S}_1$ and $\mathcal{S}_2$ are finite.

### *A. System Models and Definitions*

Consider the two classes of Helper problems with equivocations constraints shown in Figure 2. For brevity, we provide full formal definitions for the model of Figure 2a; and only outline the differences for the model of Figure 2b.

The lossy Helper problem with equivocation constraint of Figure 2a is defined by a source alphabet $(\mathcal{S}_0, \mathcal{S}_1)$, a joint input p.m.f. $P_{S_0,S_1}$, a reconstruction alphabet $\hat{\mathcal{S}}_1$ and a distortion measure defined as:

$$\begin{aligned} d_1 \ : \ \mathcal{S}_1 \times \hat{\mathcal{S}}_1 \ &\to \ [0 : \infty) \\ (s_1, \hat{s}_1) \ &\to \ d_1(s_1, \hat{s}_1) \ . \end{aligned} \tag{1}$$

**Definition 1.** *An $(n, M_n, M_{1,n})$ code for the Helper model with equivocation constraint of Figure 2a consists in:*

i) *Two message sets $\mathcal{M}_n = [1 : M_n]$ and $\mathcal{M}_{1,n} = [1 : M_{1,n}]$.*

ii) *Two encoding functions $f$ and $f_1$ defined as:*

$$\begin{aligned} f \ : \ \mathcal{S}_0^n \times \mathcal{S}_1^n \ &\to \ [1 : M_n] \\ (S_0^n, S_1^n) \ &\to \ W = f(S_0^n, S_1^n) \end{aligned} \tag{2}$$

*and*

$$\begin{aligned} f_1 \ : \ \mathcal{S}_0^n \times \mathcal{S}_1^n \ &\to \ [1 : M_{1,n}] \\ (S_0^n, S_1^n) \ &\to \ W_1 = f_1(S_0^n, S_1^n) \ . \end{aligned} \tag{3}$$

iii) *A decoding functions g defined as:*

$$g \; : \; [1:M_n]\times[1:M_{1,n}] \;\; \rightarrow \;\; \hat{\mathcal{S}}_1^n \qquad (W,W_1) \;\; \rightarrow \;\; \hat{S}_1^n \; . \tag{4}$$

*The average distortion and equivocation of a code are defined respectively as*

$$d_1^{(n)}(S_1^n,\hat{S}_1^n) = \frac{1}{n}\sum_{i=1}^{n} d_1(S_{1,i},\hat{S}_{1,i}) \quad \textit{and} \quad \frac{1}{n}H(S_0^n,S_1^n|W). \tag{5}$$

*A rate-distortion-equivocation quadruple* $(R,R_1,D_1,\Delta)$ *is said to be achievable if there exists a sequence of* $(n,M_n,M_{1,n})$ *codes such that*

$$\limsup_{n\to\infty} \log_2(M_n) \leq R \; , \tag{5a}$$

$$\limsup_{n\to\infty} \log_2(M_{1,n}) \leq R_1 \; , \tag{5b}$$

$$\limsup_{n\to\infty} \mathbb{E}\left(d_1^{(n)}(S_1^n,\hat{S}_1^n)\right) \leq D_1 \; , \tag{5c}$$

$$\liminf_{n\to\infty} \frac{1}{n}H(S_0^n,S_1^n|W) \geq \Delta \; . \tag{5d}$$

*The rate-distortion-equivocation region is the set of all achievable quadruples* $(R,R_1,D_1,\Delta)$.

**Definition 2.** *An* $(n,M_n,M_{1,n})$ *code for the Helper model with equivocation constraint of Figure 2b is defined similarly, but with the decoding function g at the legitimate receiver being a mapping*

$$g \; : \; [1:M_n]\times[1:M_{1,n}] \;\; \rightarrow \;\; \hat{\mathcal{S}}_0^n\times\hat{\mathcal{S}}_1^n \qquad (W,W_1) \;\; \rightarrow \;\; (\hat{S}_0^n,\hat{S}_1^n) \; . \tag{6}$$

*The average distortion and equivocation level achieved by such a code are given respectively by*

$$d_1^{(n)}(S_1^n,\hat{S}_1^n) = \frac{1}{n}\sum_{i=1}^{n} d_1(S_{1,i},\hat{S}_{1,i}) \quad \textit{and} \quad \frac{1}{n}H(S_0^n|W). \tag{7}$$

*A rate-distortion-equivocation quadruple* $(R,R_1,D,\Delta)$ *is said to be achievable if, similarly,* (5a)-(5c) *are satisfied, along with a distinct equivocation constraint*

$$\liminf_{n\to\infty} \frac{1}{n}H(S_0^n|W) \geq \Delta \tag{8}$$

*and where the probability of error, defined as*

$$P_e^{(n)} \triangleq \mathbb{P}\left(\hat{S}_0^n \neq S_0^n\right), \tag{9}$$

*is such that*

$$\limsup_{n\to\infty} P_e^{(n)} = 0. \tag{10}$$

*B. Helper Model with Equivocation on Both Sources*

Consider the Helper problem with equivocation constraint imposed on both sources of Figure 2a. In what follows, we provide a single-letter characterization of the optimal rate-distortion-equivocation region of this model, discuss the role of the private link and illustrate the result through a binary example.

The following theorem states the optimal rate-distortion-equivocation region of the model of Figure 2a.

**Theorem 1.** *A rate-distortion-equivocation quadruple $(R, R_1, D_1, \Delta)$ is achievable for the Helper model with equivocation constraint of Figure 2a if and only if*

$$R_1 \geq I(S_0, S_1; \hat{S}_1|U) \tag{11a}$$

$$R \geq I(U; S_0, S_1) \tag{11b}$$

$$\Delta \leq H(S_0, S_1|U) \ , \tag{11c}$$

*for some joint probability mass function $P_{US_0S_1\hat{S}_1}$ that satisfies $\mathbb{E}d_1(S_1, \hat{S}_1) \leq D_1$ .*

*Proof.* A detailed proof of the direct part, as well as of the converse part, of Theorem 1 are reported in Appendix A. Hereafter, we provide a brief outline of the achievability proof. The encoder finds a description $U^n$ that is strongly jointly typical with the pair $(S_0^n, S_1^n)$ and transmits it on the public link. It can do so as long as $n$ is large and $R \geq I(U; S_0, S_1)$. Then, accounting for that this description $U^n$ is available as a side information sequence at both encoder and decoder, it finds another description $\hat{S}_1^n$ (superimposed on $U^n$) that is strongly jointly typical with $(S_0^n, S_1^n)$ conditionally on $U^n$, and transmits it on the private link. Again, it can do so as long as $n$ is large and $R_1 \geq I(\hat{S}_1; S_0, S_1|U)$. The analysis of the equivocation of this scheme, detailed in Appendix A, shows that it is given by the conditional entropy $H(S_0, S_1|U)$. □

The following corollary specializes the result of Theorem 1 to the case of lossless reconstruction of the source $S_1^n$ at the legitimate receiver.

**Corollary 1.** *In the case of lossless reconstruction of the source $S_1^n$, i.e., $D_1 = 0$, a rate-equivocation triple $(R, R_1, \Delta)$ is achievable for the Helper model with equivocation constraint of Figure 2a if and only if*

$$R + R_1 \geq H(S_1) \tag{12a}$$

$$\Delta \leq H(S_0|S_1) + \min\{ R_1, \ H(S_1)\} \ . \tag{12b}$$

*Proof.* The proof of Corollary 1 appears in Appendix B. □

**Remark 1.** *Although the legitimate receiver requires only the source component $S_1^n$ in the model of Figure 2a, and the source component $S_0^n$ is to be kept secret (to some level) from the eavesdropper, the description $U^n$ that is sent on the public link depends not only on $S_1^n$ but also on $S_0^n$ (the a.r.v. $U$ is arbitrarily correlated with $(S_0, S_1)$ in Theorem 1). This aspect, which appears here only due to the imposed equivocation constraint, is reminiscent of Yamamoto's result [16] that a strictly smaller equivocation level is achieved if the encoder does not observe the source $S_2^n$ to keep secret in the model of Figure 1.* □

**Remark 2.** *Investigating the proof of Corollary 1 in Appendix B it can be noted that the optimal choice of the auxiliary random variable $U$ in Theorem 1 satisfies the Markov chain $U -\!\!\circ\!\!- S_1 -\!\!\circ\!\!- S_0$. This means that by opposition to the lossy case, encoder side information $S_0^n$ does not enable strictly higher equivocation levels at the eavesdropper in the case of lossless reconstruction of the source $S_1^n$.* □

**Remark 3.** *The model of Figure 1, in the case in which $S_2 = (S_0, S_1)$, can be obtained as a special case of that of Figure 2a by setting $R_1 = 0$. Thus, in this case, the result of Theorem 1 generalizes that of [16, Theorem 1] to the case in which the legitimate receiver also has a noise-free rate-limited private link.* □

**Remark 4.** *In [26], Tandon et al. study a model in which an encoder transmits a memoryless source $X^n$ to the legitimate receiver losslessly, over a noise-free rate-limited public link on which listens an external passive eavesdropper. The receiver is assisted with a rate-limited helper that observes an arbitrarily correlated source $Y^n$. They characterize the optimal rate-equivocation region of this model. In the case in which the encoder and helper observe the same source, our result of Theorem 1 generalizes that of [26] to the lossy reproduction setting.* □

The following corollary specializes the result of Theorem 1 to the case of a single source, i.e., $S_0 = S_1 = S$.

**Corollary 2** (Single Source)**.** *In the case of one source, i.e., $S_0 = S_1 = S$ in the model of Figure 2a, the rate-distortion-equivocation region is given by the set of quadruples $(R, R_1, D, \Delta)$ such that*

$$R_1 \geq I(S; \hat{S}|U) \tag{13a}$$

$$R \geq I(U; S) \tag{13b}$$

$$\Delta \leq H(S|U) \tag{13c}$$

*for some joint distribution $P_{US\hat{S}}$ that satisfies $\mathbb{E}d_1(S, \hat{S}) \leq D_1$.*

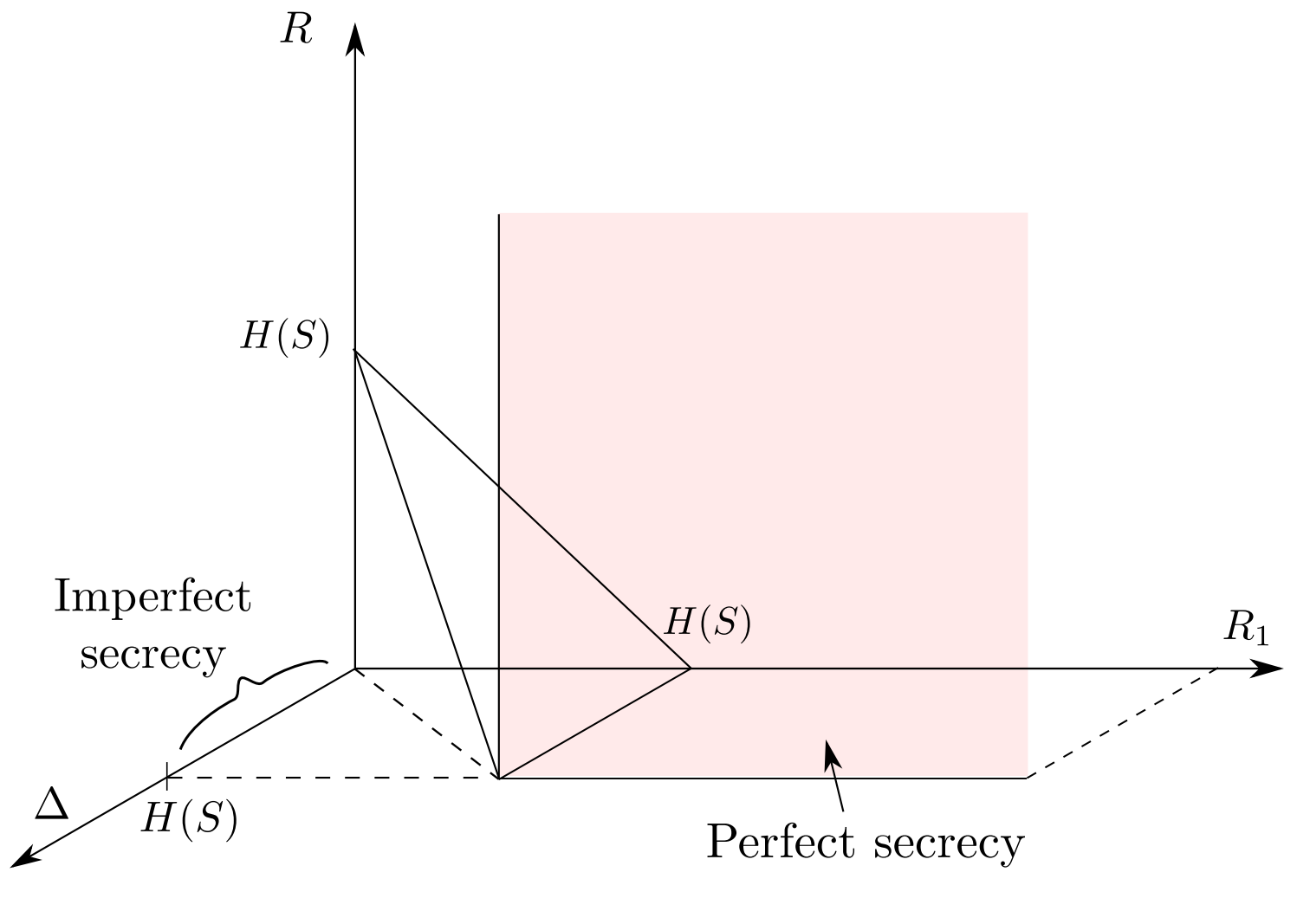

Fig. 4: Rate-equivocation region for the secure lossless Helper problem

In the case of lossless reconstruction of the source $S^n$, the rate-equivocation region of Corollary 2 can also be obtained from Corollary 1 by setting $S_0 = S_1 = S$ therein. As expected in this case, in order to guarantee full equivocation at the eavesdropper, i.e., $\Delta = H(S)$, information has to transit through the private link which then should have capacity no smaller than $H(S)$ – the public link can carry independent information, but not any useful information about $S$. (A useful connection with secret-sharing and encryption based schemes, in which even the public link can carry an encrypted version of the source, will be made in the next section). In the case in which $R_1 < H(S)$, perfect secrecy is not possible, and the equivocation at the eavesdropper is given by the amount of information that it does not intercept, i.e., $R_1$ precisely.

### *C. Helper Model with Reproduction of Both Sources*

Consider the Helper problem of Figure 2b, with the equivocation constraint $\Delta$ imposed only on the source component $S_0^n$ and the legitimate receiver reproducing the source $S_0^n$ losslessly and the source $S_1^n$ lossily, to within average distortion $D_1$. As it will become apparent from the discussion that will follow, the analysis of this model is less easy than that of the model of Figure 2a in which the legitimate receiver reproduces only the source component $S_1^n$.

*1) Rate-Distortion-Equivocation Region:* The following theorem states the optimal rate-distortion-equivocation region of the Helper model with equivocation constraint of Figure 2b.

**Theorem 2.** *A rate-distortion-equivocation triple* $(R, R_1, D_1)$ *is achievable for the Helper model with equivocation constraint of Figure 2b if and only if*

$$R + R_1 \geq H(S_0) + \min_{P_{\hat{S}_1|S_0,S_1}} I(\hat{S}_1; S_1|S_0) \; , \tag{14a}$$

$$\Delta \leq \min\{H(S_0), R_1\} \; , \tag{14b}$$

*for some conditional* $P_{\hat{S}_1|S_0,S_1}$ *that satisfies* $\mathbb{E}d_1(S_1, \hat{S}_1) \leq D_1$ .

*Proof.* Detailed proofs of the direct and converse parts of Theorem 2 appear in Appendix C. Hereafter we provide a brief outline of the proof of achievability. The encoder uses both private and public links to describe the source $S_0^n$ losslessly to the legitimate receiver. In doing so, the rate splitting of the required $H(S_0)$ bits per sample among the two links is dictated by desired level of equivocation $\Delta$ about the source $S_0^n$ at the eavesdropper. Then, conditioned on $S_0^n$, the encoder describes the source $S_1^n$ through both private and common links, accounting for $S_0^n$ as a side information sequence that is available at both encoder and decoder. More specifically, the coding scheme that we use for the proof of Theorem 2 uses a careful combination of binning and rate-splitting. Let $R_1 = R_{1,1} + R_{1,0}$ and $R = R_{0,0} + R_{0,1}$. Assign the set of typical $S_0^n$'s to $2^{n(R_{0,0}+R_{1,0})}$ bins indexed with $w_{0,0} = 1, \ldots, 2^{nR_{0,0}}$ and $w_{1,0} = 1, \ldots, 2^{nR_{1,0}}$. For every typical $S_0^n$, generate $2^{n(R_{1,1}+R_{0,1})}$ independent sequences $\hat{S}_1^n(w_{1,1}, w_{0,1})$, each i.i.d. according to the product of $P_{\hat{S}_1|S_0}$, indexed with $w_{1,1} = 1, \ldots, 2^{nR_{1,1}}$ and $w_{0,1} = 1, \ldots, 2^{nR_{0,1}}$. For every typical pair $(S_0^n, S_1^n)$, the encoder finds the indices $(w_{0,0}, w_{1,0})$ of the bin in which lies $S_0^n$, and then finds a pair $(w_{1,1}, w_{0,1})$ such that $\hat{S}^n(w_{1,1}, w_{0,1})$ is strongly jointly typical with $S_1^n$ conditionally on $S_0^n$. It can do so as long as $n$ is large, $R_{0,0} + R_{1,0} \geq H(S_0)$ and $R_{1,1} + R_{0,1} \geq I(\hat{S}_1; S_1|S_0)$. Next, the encoder transmits the indices $w_{0,0}$ and $w_{0,1}$ on the public link and the indices $w_{1,0}$ and $w_{1,1}$ on the private link. A block diagram of the indices and their transmission on the public and private links is shown in Figure 5. The legitimate receiver first recovers $S_0^n$ and then reconstructs the description of $S_1^n$ as $\hat{S}_1^n(w_{1,1}, w_{0,1})$. Recalling that the source $S_1^n$ is arbitrarily correlated with the source $S_0^n$ which to be kept secret to within equivocation level $\Delta$ at the eavesdropper, the analysis of the equivocation of this scheme requires showing that the compression index $w_{0,1}$ of $S_1^n$ that is sent on the public link leaks no additional information about $S_0^n$ beyond what the transmission of the index $w_{0,0}$ of $S_0^n$ on the public link leaks about this source, which we prove through a here established new counting lemma (see Lemma 1 in Appendix C or a different proof of a similar result in [31]).

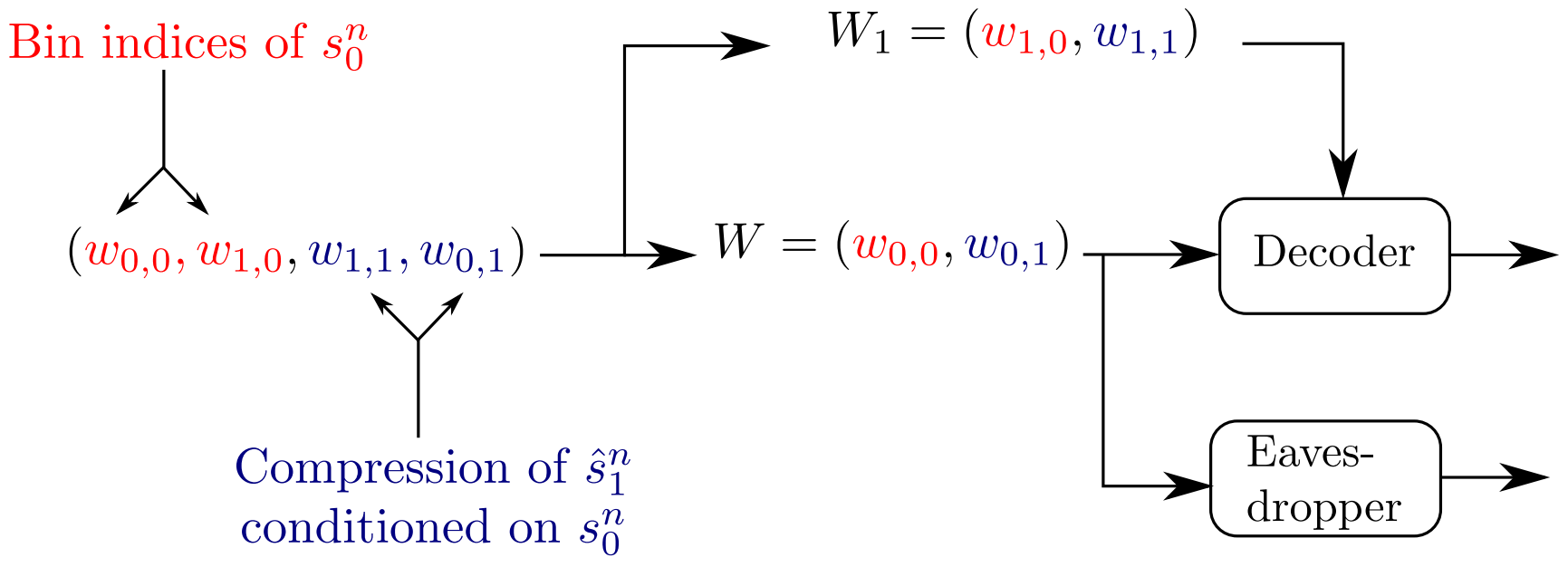

Fig. 5: Block diagram of the compression indices and transmission on the public and private links for the model of Figure 2b.

□

The following corollary specializes the result of Theorem 2 to the case of maximum equivocation at the eavesdropper, i.e., $\Delta = H(S_0)$.

**Corollary 3** (Case of Perfect Secrecy). *For the Helper model of Figure 2b, a rate-distortion triple* $(R, R_1, D_1)$ *is achievable with perfect secrecy if and only if*

$$R_1 \geq H(S_0) \tag{15a}$$

$$R + R_1 \geq H(S_0) + \min_{P_{\hat{S}_1|S_0S_1}} I(\hat{S}_1; S_1|S_0) \tag{15b}$$

*for some conditional* $P_{\hat{S}_1|S_0S_1}$ *that satisfies* $\mathbb{E}d_1(S_1, \hat{S}_1) \leq D_1$ .

The following remarks help understanding better the result of Theorem 2. They also provide some useful connections with secret-sharing and encryption-based schemes, as well as results from related previous work.

**Remark 5.** *In the case of a single source* $S_0^n$ *that is to be conveyed losslessly to the legitimate receiver and kept secret from the eavesdropper, i.e.,* $S_1 = \emptyset$ *or, equivalently,* $D_1 = \infty$*, the result of Theorem 2 reduces, under perfect secrecy ,i.e.* $\Delta = H(S_0)$*, to* $R_1 \geq H(S_0)$*, which is consistent with the observation that, in this case, secure transmission is possible only through the private link which then should have its capacity no smaller than* $H(S_0)$ *in order for the encoder to describe* $S_0^n$ *losslessly to the legitimate receiver. The result in this specific case can also be inferred from [26] and [27]. Also, without the secrecy constraint, i.e.* $\Delta = 0$*, the model of*

*Figure 2b reduces to a multiple descriptions problem with combined reconstruction only [32, page 323]; and the rate region, which is symmetric in $R$ and $R_1$ in this case, is described by the single constraint* (15b). □

**Remark 6.** *By opposition to the model of Figure 2a in which the constraint of perfect secrecy precludes the usage of the public link, for the model of Figure 2b this link can still be utilized, e.g., to transmit a description of the part of $S_1^n$ that is not correlated with $S_0^n$ (i.e., the innovation of $S_1^n$ relative to $S_0^n$ or $S_1^n$ conditionally on $S_0^n$). Recalling that the associated compression index can be shown to be independent of the source $S_0^n$ (see Lemma 1 in Appendix C, or an alternate proof in [31, Section III]), this leaks no additional information about $S_0^n$ to the eavesdropper as we already mentioned in the outline proof of achievability.* □

**Remark 7.** *In the above outlined coding scheme of Theorem 2, the uncertainty at the eavesdropper is induced only through appropriate binning. For instance, no encryption is utilized. The reader may wonder whether alternate coding schemes that rely on encryption and/or secret-sharing can be employed, since some common randomness can be shared through the private link and then utilized for the transmission on the public link. In the next section we develop a coding scheme that is equally optimal and is based on secret-sharing and encryption. For example, in the case of perfect secrecy, i.e., $\Delta = H(S_0)$, for coding schemes without encryption or secret sharing the constraint $R_1 \geq H(S_0)$ of the converse part of Corollary 3 simply means that the source $S_0^n$ that is to be kept secret should be sent entirely only through the private link (so as not to be intercepted by the eavesdropper). Alternatively, for coding schemes that are based on encryption or secret sharing, this constraint expresses the fact that, for the source $S_0^n$ to be transmitted directly over the public link without leaking any information about it to the eavesdropper, the key utilized for its encryption or that shared through secret-sharing approaches should be at least $H(S_0)$ bits long.* □

*2) Alternate Coding Scheme: Secret-Sharing and Encryption:*

In this section, we develop an alternate optimal coding scheme for the Helper model with equivocation constraint of Figure 2b. By opposition to the scheme of Theorem 2, this scheme utilizes the private link as a means of sharing secret between the transmitter and the legitimate receiver, in the form of all or part of the source $S_1^n$. The shared secret is then utilized as a key to encrypt all or part of the source $S_0^n$, depending on the desired level of equivocation and transmit it directly over the public link. The result is stated in the following theorem.

**Theorem 3.** *For the Helper model with equivocation constraint of Figure 2b, there is an alternate coding scheme that utilizes secret-sharing and encryption which is optimal, i.e., achieves the rate-distortion-equivocation-region of Theorem 2.*

*Proof.* A formal proof of Theorem 3 is given in Appendix E. The coding scheme is similar to that of Theorem 2; and so, hereafter, we only outline the main steps of it. Recall the rate splitting and the binning used in the coding scheme of Theorem 2. Also, recall the following important two elements therein: i) the index $w_{1,1}$, which together with $w_{0,1}$ identify the sequence $\hat{S}_1^n(w_{1,1}, w_{0,1})$ that is found jointly typical with $S_1^n$ conditionally on $S_0^n$, is transmitted over the private link; and ii) the index $w_{1,0}$, which together with $w_{0,0}$ identify the bin in which lies the source $S_0^n$, is transmitted over the private link. In the coding scheme that we use to show Theorem 3, the legitimate transmitter-receiver pair use the index $w_{1,1}$ of the description of the source $S_1^n$ as a secret-key. Specifically, the encoder encrypts the message $w_{1,0}$ by taking the bit-wise modulo-2 sum of the binary expansion of $w_{1,0}$ and $w_{1,1}$; and transmits it over the public link. If the rate of the message $w_{1,1}$ that is utilized as a secret key does not suffice to encrypt all of $w_{1,0}$, only part of the latter is encrypted. □

*3) Binary Example:* Consider the Helper model with equivocation constraint of Figure 2b. Let $(S_0, S_1) \sim$ DSBS($p$) for some $p \in [0, 1]$, i.e., $\mathcal{S}_0 = \mathcal{S}_1 = \{0, 1\}$ and for $(s_0, s_1) \in \{0, 1\}^2$,

$$P_{S_0,S_1}(s_0, s_1) = \frac{1-p}{2}\delta(s_0, s_1) + \frac{p}{2}(1 - \delta(s_0, s_1)). \tag{16}$$

Also, assume a Hamming distortion measure $d_H$. In this case, it is easy to see that

$$\min_{P_{\hat{S}_1|S_0S_1}} I(\hat{S}_1; S_1|S_0) = [h_2(p) - h_2(D_1)]^+ , \tag{17}$$

where the minimum is achieved with the choice of the test channel $\hat{S}_1 = S_1 \oplus S_0 \oplus U$ and $U$ is a Bernoulli random variable with parameter $\left(\frac{p-2D_1}{2p-1}\right)$. Thus, for this binary example, Theorem 2 reduces to the set of quadruples $(R, R_1, D_1, \Delta)$ that satisfy

$$\mathcal{R}_{\text{secret}} : \begin{cases} R + R_1 & \geq 1 + [h_2(p) - h_2(D_1)]^+ \\ \Delta & \leq \min\{1, R_1\} \end{cases} \tag{18}$$

This region is shown in Figure 6 where we project over all possible equivocation levels. For comparison reasons, the figure also shows the rate-distortion region of this example if there were no secrecy constraint, i.e., the region defined by

$$\mathcal{R}_0 \ : \ R + R_1 \geq 1 + [h_2(p) - h_2(D_1)]^+ . \tag{19}$$

The shaded region in the figure delimits the region in which perfectly secure transmission is not possible.

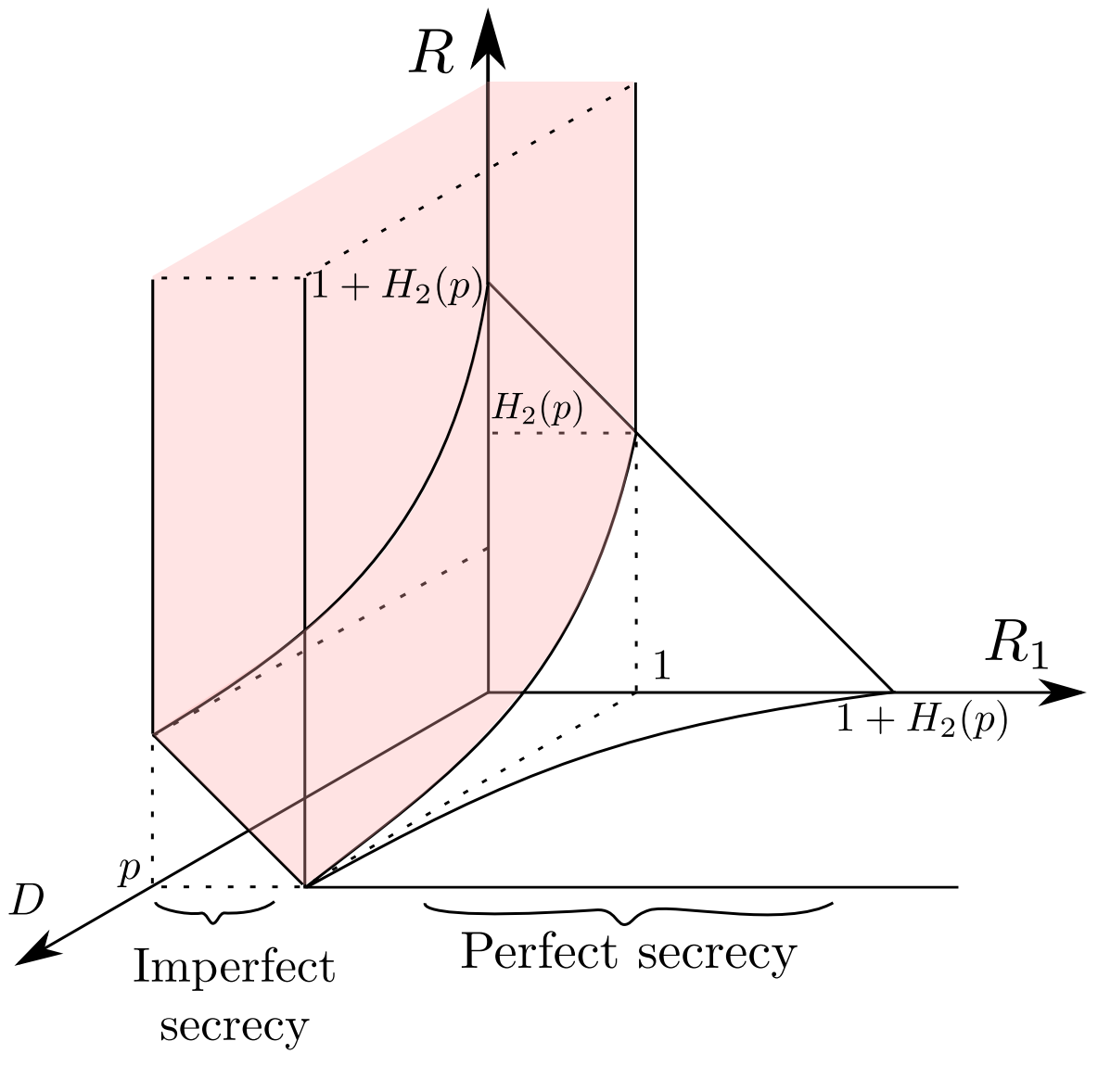

Fig. 6: Secure rate-distortion region of the Helper problem of Figure 2b. Case of DSBS(p) sources

### D. Digression

Consider again the Helper model with equivocation constraint of Figure 2. A more general setting, not treated in this paper, is one in which the role of the encoder is to convey (only) the source component $S_1^n$ to the legitimate receiver lossily, i.e., to within some desired fidelity level $D_1$, while maintaining the equivocation about (only) the source component $S_0^n$ at the eavesdropper no smaller than some prescribed level $\Delta$. This problem was solved by Yamamoto in [16] in the specific case in which there is no private link to the legitimate receiver, i.e., $R_1 = 0$. However, solving it in the general case seems to hinge upon characterizing the *common information* [33], [34] of the two dependent random variables $S_0$ and $S_1$. To see this, consider for example the case of perfect secrecy and lossless reconstruction, i.e., $D_1 = 0$ and $\Delta = H(S_0)$. In this case, intuitively, every information that $S_1$ can give about $S_0$ should *not* be transmitted over the public link, so as not to be intercepted by the eavesdropper.

## III. Secure Lossy Gray-Wyner Problems

In this section, we study the Gray-Wyner model of Figure 3. It is assumed that the alphabets $\mathcal{S}_0$, $\mathcal{S}_1$ and $\mathcal{S}_2$ are finite.

### A. System Models and Definitions

Consider the two classes of Gray-Wyner problems with equivocations constraints shown in Figure 3. For convenience, first we provide full formal definitions for the model of Figure 3b; and then only highlight the differences for the model of Figure 3a.

The lossy Gray-Wyner problem of Figure 3b is defined by a product source alphabet $\mathcal{S}_0 \times \mathcal{S}_1 \times \mathcal{S}_2$, a joint input p.m.f. $P_{S_0,S_1,S_2}$, a reconstruction product alphabet $\hat{\mathcal{S}}_1 \times \hat{\mathcal{S}}_2$ and two distortion measures defined, for $j = 1,2$, as

$$\begin{aligned} d_j \ : \ \mathcal{S}_j \times \hat{\mathcal{S}}_j &\rightarrow [0:\infty) \\ (s_j, \hat{s}_j) &\rightarrow d_j(s_j, \hat{s}_j) \ . \end{aligned} \tag{20}$$

**Definition 3.** *An $(n, M_n, M_{1,n}, M_{2,n})$ code for the lossy Gray-Wyner problem with equivocation constraint of Figure 3b consists of:*

i) *Three message sets $\mathcal{M}_n = [1 : M_n]$, $\mathcal{M}_{1,n} = [1 : M_{1,n}]$ and $\mathcal{M}_{2,n} = [1 : M_{2,n}]$*
ii) *Three encoding functions $f$, $f_1$ and $f_2$ defined as*

$$\begin{aligned} f \ : \ \mathcal{S}_0^n \times \mathcal{S}_1^n \times \mathcal{S}_2^n &\rightarrow [1 : M_n] \\ (S_0^n, S_1^n, S_2^n) &\rightarrow W = f(S_0^n, S_1^n, S_2^n) \end{aligned} \tag{21}$$

*and, for $j = 1,2$,*

$$\begin{aligned} f_j \ : \ \mathcal{S}_0^n \times \mathcal{S}_1^n \times \mathcal{S}_2^n &\rightarrow [1 : M_{j,n}] \\ (S_0^n, S_1^n, S_2^n) &\rightarrow W_j = f_j(S_0^n, S_1^n, S_2^n) \ . \end{aligned} \tag{22}$$

iii) *Two decoding functions $g_1$ and $g_2$ defined, for $j = 1,2$, as*

$$\begin{aligned} g_j \ : \ [1 : M_n] \times [1 : M_{j,n}] &\rightarrow \hat{\mathcal{S}}_{0,j}^n \times \hat{\mathcal{S}}_j^n \\ (W, W_j) &\rightarrow (\hat{S}_{0,j}^n, \hat{S}_j^n) \ . \end{aligned} \tag{23}$$

*The average distortions and equivocation achieved by such a code are given by*

$$d_1^{(n)}(S_1^n, \hat{S}_1^n) \ , \quad d_2^{(n)}(S_2^n, \hat{S}_2^n) \quad \textit{and} \quad \frac{1}{n}H(S_0^n|W) \ ; \tag{24}$$

*and the probability of error about the source $S_0^n$ is defined as*

$$P_e^{(n)} \triangleq \mathbb{P}\Big(\hat{S}_{0,1}^n \neq S_0^n \ \ \textit{or} \ \ \hat{S}_{0,2}^n \neq S_0^n\Big). \tag{25}$$

*A rate-distortion-equivocation tuple $(R, R_1, R_2, D_1, D_2, \Delta)$ is said to be achievable for the model of Figure 3b if there exists a sequence of codes $(n, M_n, M_{1,n}, M_{2,n})$ such that*

$$\limsup_{n\to\infty} \frac{1}{n} \log_2(M_n) \leq R \ , \tag{26a}$$

$$\limsup_{n\to\infty} \frac{1}{n} \log_2(M_{j,n}) \leq R_j \quad \textit{for} \quad j = 1, 2 \ , \tag{26b}$$

$$\limsup_{n\to\infty} \mathbb{E}\left(d_j^{(n)}(S_j^n, \hat{S}_j^n)\right) \leq D_j \quad \textit{for} \quad j = 1, 2 \ , \tag{26c}$$

$$\limsup_{n\to\infty} P_e^{(n)} = 0 \ , \tag{26d}$$

$$\liminf_{n\to\infty} \frac{1}{n}H(S_0^n|W) \geq \Delta. \tag{26e}$$

*The rate-distortion-equivocation region for the model of Figure 3b is the set of all achievable tuples $(R, R_1, R_2, D_1, D_2, \Delta)$.*

□

**Definition 4.** *For the model of Figure 3a the definition of an $(n, M_n, M_{1,n}, M_{2,n})$ code is similar to in definition 3 and is obtained by setting $S_0 = \emptyset$ therein. But, the average distortions and equivocation achieved by such a code are given by*

$$d_1^{(n)}(S_1^n, \hat{S}_1^n) \ , \quad d_2^{(n)}(S_2^n, \hat{S}_2^n) \quad \textit{and} \quad \frac{1}{n}H(S_1^n, S_2^n|W) \ . \tag{27}$$

*A rate-distortion-equivocation array $(R, R_1, R_2, D_1, D_2, \Delta)$ is said to be achievable for the model of Figure 3a if there exists a sequence of codes $(n, M_n, M_{1,n}, M_{2,n})$ such that*

$$\limsup_{n\to\infty} \frac{1}{n} \log_2(M_n) \leq R \ , \tag{28a}$$

$$\limsup_{n\to\infty} \frac{1}{n} \log_2(M_{j,n}) \leq R_j \quad \textit{for} \quad j = 1, 2 \ , \tag{28b}$$

$$\limsup_{n\to\infty} \frac{1}{n} \mathbb{E} d_j^{(n)}(S_j^n, \hat{S}_j^n) \leq D_j \quad \textit{for} \quad j = 1, 2 \ , \tag{28c}$$

$$\liminf_{n\to\infty} \frac{1}{n}H(S_1^n, S_2^n|W) \geq \Delta. \tag{28d}$$

*The rate-distortion-equivocation region for the model of Figure 3a is the set of all achievable tuples $(R, R_1, R_2, D_1, D_2, \Delta)$.*

### *B. Gray-Wyner Model with Equivocation on Both Sources*

Consider the lossy Gray-Wyner model with equivocation constraint imposed on both sources of Figure 3a. In what follows, we provide a single-letter characterization of the region of optimal tradeoffs among rate triples $(R, R_1, R_2)$, average distortion pairs $(D_1, D_2)$ and equivocation level $\Delta$. Also, we discuss the implications of the imposed secrecy constraint on Gray-Wyner's original network [18], and illustrate the result through a binary example.

*1) Rate-Distortion-Equivocation Region:* The following theorem states the optimal rate-distortion-equivocation region of the Gray-Wyner model of Figure 3a. .

**Theorem 4.** *A rate-equivocation-distortion tuple $(R, R_1, R_2, D_1, D_2, \Delta)$ is achievable for the Gray-Wyner model with equivocation constraint of Figure 3a if and only if*

$$R \geq I(U; S_1, S_2) \ , \tag{29a}$$

$$R_1 \geq I(\hat{S}_1; S_1, S_2|U) \ , \tag{29b}$$

$$R_2 \geq I(\hat{S}_2; S_1, S_2|U) \ , \tag{29c}$$

$$\Delta \leq H(S_1, S_2|U) \ , \tag{29d}$$

*for some joint distribution $P_{U,S_1,S_2,\hat{S}_1,\hat{S}_2}$ that satisfies $\mathbb{E}d_1(S_1, \hat{S}_1) \leq D_1$ and $\mathbb{E}d_2(S_2, \hat{S}_2) \leq D_2$.*

*Proof.* A detailed proof of Theorem 4 appears in Appendix F. □ □

**Remark 8.** *It is insightful to observe that the imposed secrecy constraint changes drastically the usage of the common and private links by the encoder in the Gray-Wyner model without secrecy constraints. To see this, consider for example the case of lossless reconstruction and perfect secrecy, i.e., $D_1 = D_2 = 0$ and $\Delta = H(S_1, S_2)$. In this case, the encoder should send no information on the common link, even if the two sources are correlated and/or exhibit some "common information". In the extreme case in which $S_1 = S_2 = S$, a full description of $S$ is sent twice, on the two private links, even if this entails some redundancy which could be saved had there been no secrecy constraint.* □

*2) Binary Example:*

Consider the Gray-Wyner model with secrecy constraints of Figure 3a. Let $(S_1, S_2) \sim$ DSBS($p$), for some $p \in [0, \frac{1}{2}]$. For simplicity, we focus on the lossless compression case, i.e., $D_1 = D_2 = 0$, and symmetric rates on the private links, i.e., $R_1 = R_2$.

Without an equivocation constraint, i.e. $\Delta = 0$, the optimal rate region is obtained by specializing the result of [18, Theorem 1] to this binary example. However, the optimal choice of the auxiliary random variable involved in [18, Theorem 1] is still not known in general for the binary sources case; and the best explicit inner bound on the region $(R, R_1)$, also provided in [18], is shown in Figure 7. It is represented by the segments $[AF]$, $[FG]$ and $[GB]$, where the corner points $A, B, F$ and $G$ are the points with coordinates

$$A \ : \ (1 + h_2(p), 0, 0) \tag{30a}$$

$$B \ : \ (0, 1, 0) \tag{30b}$$

$$F \ : \ \left(1, \frac{h_2(p)}{2}, 0\right) \tag{30c}$$

$$G \ : \ \left(1 + h_2(p) - 2h_2\left(\frac{1 - \sqrt{1-2p}}{2}\right), h_2\left(\frac{1 - \sqrt{1-2p}}{2}\right), 0\right), \tag{30d}$$

and correspond respectively to the following choices of the auxiliary random variable $U$, $U = U_A \triangleq (S_1, S_2)$, $U = U_B \triangleq \emptyset$, $U = U_F$ is obtained through time-sharing between $U = S_1$ and $U = S_2$, and $U = U_G \triangleq S_1 \oplus U_0 = S_2 \oplus U_0$ where $U_0 \sim \text{Bern}(p_1)$, with $p_1 \in [0, 1/2]$ chosen such that $p = 2p_1 \cdot (1 - p_1)$, i.e., $p_1 = 1/2 - (\sqrt{1-2p})/2$. The figure also shows an outer bound on the region $(R, R_1)$ represented by the segments $[AH]$ and $[HB]$ that correspond to the intersection of the two constraints (Pangloss planes):

$$R + R_1 \geq 1 \tag{31a}$$

$$R + 2R_1 \geq 1 + h_2(p) \; . \tag{31b}$$

Conversely, in the case of *perfect secrecy*, i.e., $\Delta = H(S_1, S_2) = 1 + h_2(p)$, it is easy to see that, for this binary example and under the constraint that $R_1 = R_2$, the region of Theorem 4 reduces the set of rates satisfying

$$R_1 = R_2 \geq = 1 \; , \tag{32a}$$

$$R \geq 0 \; . \tag{32b}$$

Thus, in this case, it is optimal to transmit the source $S_1^n$ entirely on the first private link and the source $S_2^n$ entirely on the second private link; and not use the public link at all. Observe that this incurs some rate-redundancy, which is reflected, e.g., through the inequality $R + 2R_1 = 2 > H(S_1, S_2)$. Furthermore, observe that the transmit strategies described by $U_F$ and $U_G$ yield levels of equivocation given respectively by $h_2(p)$ and $2h_2(p_1)$. The associated rate-equivocation points are represented in Figure 7 respectively as the points $\tilde{F}$ and $\tilde{G}$ with coordinates

$$\tilde{F} \; : \; \left(1, \frac{h_2(p)}{2}, h_2(p)\right) \tag{33a}$$

$$\tilde{G} \; : \; \left(1 + h_2(p) - 2h_2\left(\frac{1-\sqrt{1-2p}}{2}\right), h_2\left(\frac{1-\sqrt{1-2p}}{2}\right), 2h_2\left(\frac{1-\sqrt{1-2p}}{2}\right)\right). \tag{33b}$$

### C. Gray-Wyner Model with Reproduction of Both Sources

Consider now the Gray-Wyner model with equivocation constraint of Figure 3b, in which the legitimate receivers also reproduce the source $S_0^n$ losslessly and the secrecy constraint is imposed only on $S_0^n$. As we already mentioned, the component $S_0^n$ here may represent, e.g., some common information of the sources $S_1^n$ and $S_2^n$ – the sources $S_1^n$ and $S_2^n$ need not be independent conditionally on $S_0^n$, however. Recalling that in the original Gray-Wyner network without secrecy constraints [18], the common component of the sources that are to be conveyed the two receivers should be transmitted over the common link, the imposed secrecy constraint creates a tension among reducing rate by sending $S_0^n$ on the common public link and concealing it from the eavesdropper that overhears on this link.

*1) Secure Rate-Distortion Region:* The following theorem states the optimal rate-distortion region of the Gray-Wyner model with equivocation constraint of Figure 3b in the case of perfect secrecy, i.e., $H(S_0^n|W) = nH(S_0)$.

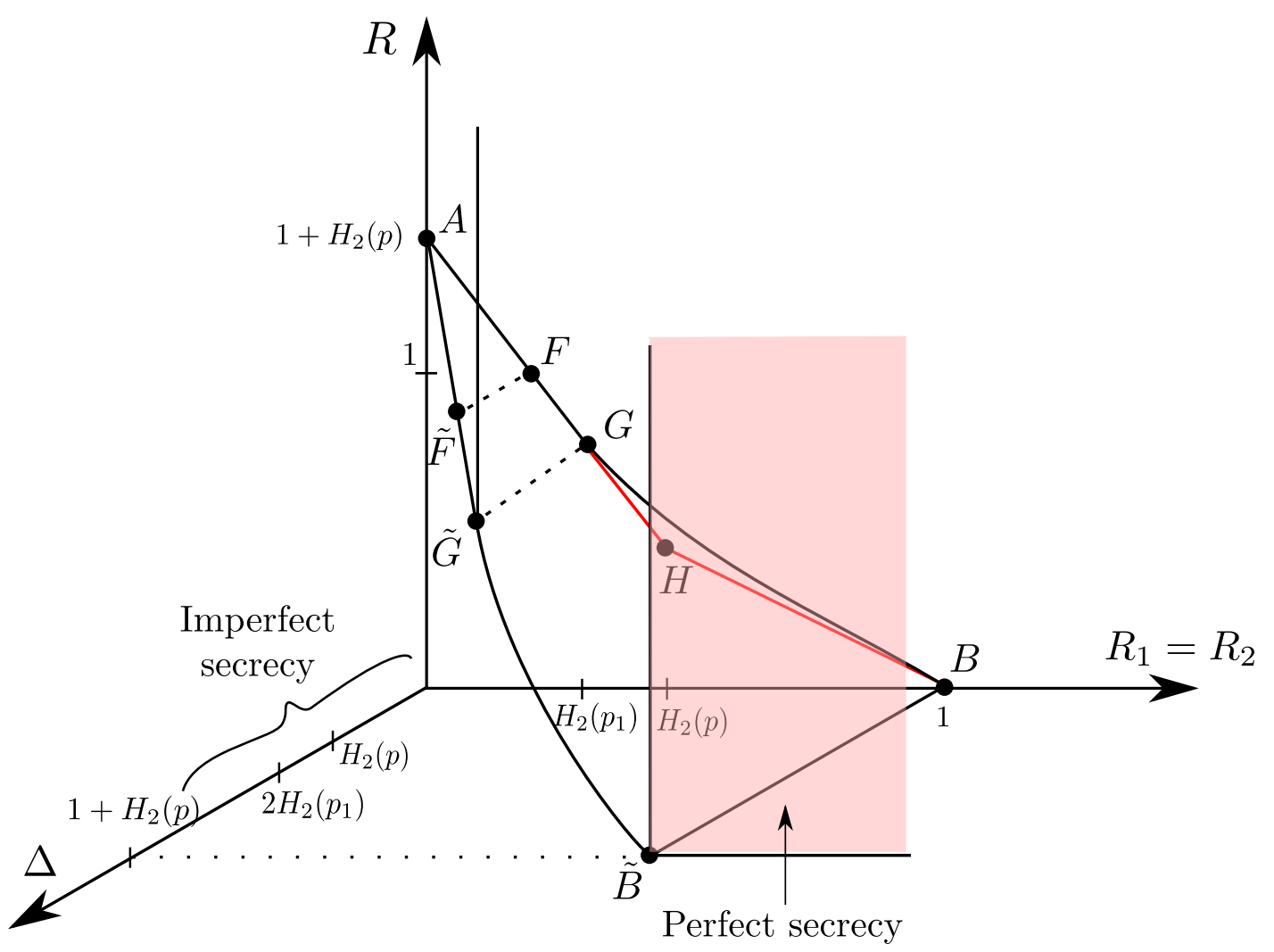

Fig. 7: Inner and outer bounds on the rate-equivocation region for the Gray-Wyner model with equivocation constraint of Figure 3a. Case of DSBS($p$) sources and symmetric rates on the private links, i.e., $R_1 = R_2$.

**Theorem 5.** *For the Gray-Wyner model with equivocation constraint of Figure 3b, a tuple* $(R, R_1, R_2, D_1, D_2, \Delta)$ *is achievable if and only if:*

$$R \geq I(U; S_1, S_2|S_0) \tag{34a}$$

$$R_1 \geq \Delta + I(\hat{S}_1; S_1, S_2|U, S_0) \tag{34b}$$

$$R_2 \geq \Delta + I(\hat{S}_2; S_1, S_2|U, S_0) \tag{34c}$$

$$R + R_1 \geq H(S_0) + I(U, \hat{S}_1; S_1, S_2|S_0) \tag{34d}$$

$$R + R_2 \geq H(S_0) + I(U, \hat{S}_2; S_1, S_2|S_0) \tag{34e}$$

$$\Delta \leq H(S_0)\ , \tag{34f}$$

*for some joint distribution* $P_{U,S_0,S_1,S_2,\hat{S}_1,\hat{S}_2}$ *that satisfies*

$$\mathbb{E}d_1(S_1, \hat{S}_1) \leq D_1 \quad \textit{and} \quad \mathbb{E}d_2(S_2, \hat{S}_2) \leq D_2. \tag{35}$$

*Proof.* The proof of Theorem 5 appears in Appendix G. Hereafter, we provide a brief outline of the proof of achievability. The encoder generates a lossless description of the source component $S_0^n$, with rate $H(S_0)$, and uses rate splitting to transmit it on both private and common links – the rate of the part of $S_0^n$ that is transmitted on the public link is chosen so as to leak $(H(S_0)-\Delta)$ bits per sample about $S_0^n$ to the eavesdropper. Superimposed on this description, the encoder generates a description $U^n$ of the pair $(S_1^n, S_2^n)$ that it transmits only on the public link. Also, superimposed on both descriptions, it sends a private description $\hat{S}_1^n$ of the pair $(S_1^n, S_2^n)$ on the private link to the first legitimate receiver, and a private description $\hat{S}_2^n$ of the pair $(S_1^n, S_2^n)$ on the private link to the second legitimate receiver.

Each legitimate receiver first recovers $S_0^n$ losslessly from both links, and then it recovers the common description $U^n$ from the common link. Finally, it recovers its dedicated description from its private link. The analysis of the equivocation of this scheme shows that, although the description $U^n$ which is sent on the public link is arbitrarily correlated with the source component to be kept secure $S_0^n$, it reveals no additional information about $S_0^n$ to the eavesdropper beyond what the part of this source that is transmitted on the public link reveals about itself. Thus, the equivocation about the source $S_0^n$ that is allowed by this scheme is precisely $\Delta$. □

The following Corollary specializes the result of Theorem 5 to the case of perfect security, i.e., $\Delta = H(S_0)$.

**Corollary 4.** *In the case of perfect secrecy, a tuple $(R, R_1, R_2, D_1, D_2)$ is achievable for the Gray-Wyner model with equivocation constraint of Figure 3b if and only if:*

$$R \geq I(U; S_1, S_2|S_0) \tag{36a}$$

$$R_1 \geq H(S_0) + I(\hat{S}_1; S_1, S_2|U, S_0) \tag{36b}$$

$$R_2 \geq H(S_0) + I(\hat{S}_2; S_1, S_2|U, S_0)\ , \tag{36c}$$

*for some joint distribution $P_{U,S_0,S_1,S_2,\hat{S}_1,\hat{S}_2}$ that satisfies*

$$\mathbb{E}d_1(S_1, \hat{S}_1) \leq D_1 \quad \text{and} \quad \mathbb{E}d_2(S_2, \hat{S}_2) \leq D_2. \tag{37}$$

**Remark 9.** *It is insightful to observe that the region of Corollary 4 can be written equivalently as the set of rate triples $(R, R_1, R_2)$ that satisfy*

$$R \geq I(U; S_1, S_2|S_0)\ , \tag{38a}$$

$$R_1 - H(S_0) \geq I(\hat{S}_1; S_1, S_2|U, S_0)\ , \tag{38b}$$

$$R_2 - H(S_0) \geq I(\hat{S}_2; S_1, S_2|U, S_0)\ , \tag{38c}$$

*for some joint $P_{U,S_0,S_1,S_2,\hat{S}_1,\hat{S}_2}$ that satisfies* (35)*. This region can be interpreted as that of a standard Gray-Wyner network without secrecy constraint [18] but with side information $S_0^n$ available at the encoder and both decoders. qed*

*2) Binary Example:* Consider the Gray-Wyner model of Figure 3b with the following choice of the sources $S_0$, $S_1$ and $S_2$. All the sources are binary valued, $(S_0, S_1) \sim$ DSBS($p$) and $(S_0, S_2) \sim$ DSBS($p$) such that $S_1 -\!\!\!\ominus\!\!\!- S_0 -\!\!\!\ominus\!\!\!- S_2$ forms a Markov chain, as shown in Figure 8. Also, let $d_1$ and $d_2$ be Hamming distortion measures. For simplicity, we concentrate on the case of lossless reconstruction of the sources $S_1$ and $S_2$ at the legitimate receivers, i.e., $D_1 = D_2$; and we investigate the symmetric rate case, i.e., $R_1 = R_2$.

Using Corollary 4, it is easy to see that the region $\mathcal{R}_{\text{secret}}$ of rate triples $(R, R_1, R_2)$ with $R_1 = R_2$ such that the first receiver reconstruct the pair $(S_0, S_1)$ losslessly, the second receiver reconstructs the pair $(S_0, S_2)$ losslessly, while the eavesdropper gets no information about the source component $S_0$ is given by

$$\mathcal{R}_{\text{secret}} : \begin{cases} R_1 = R_2 \geq 1 \ , \\ R + 2R_1 \geq 2(1 + h_2(p)) \ . \end{cases} \tag{39}$$

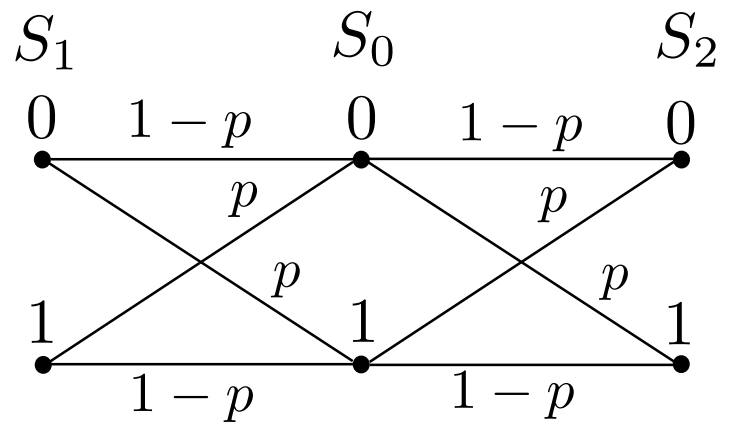

Fig. 8: Example binary sources $(S_0, S_1, S_2)$ for the Gray-Wyner model with equivocation constraint of Figure 3b.

This region is represented in Figure 9 by the region delimited by the segment $[AB]$ and the vertical line ending at $B$. The corner points $A$ and $B$ have coordinates

$$\text{A} \ : \ (1 + h_2(p) \ , \ 0) \ , \tag{40a}$$

$$\text{B} \ : \ (1 \ , \ 2h_2(p)) \ , \tag{40b}$$

and are obtained by computing the region of Corollary 4 respectively with the choices $(U, \hat{S}_1, \hat{S}_2) = (S_0, S_1, S_2)$ and $(U, \hat{S}_1, \hat{S}_2) = ((S_1, S_2), S_1, S_2)$.

For comparison, we also investigate the region of optimal triples $(R, R_1, R_2)$ for this same setting *without* secrecy constraints at all. This region can be shown easily to be given by the set of rates satisfying

$$R + R_1 \geq 1 + h_2(p) \ , \tag{41a}$$

$$R + 2R_1 \geq 1 + 2h_2(p) \ . \tag{41b}$$

It is represented in Figure 9 by the region delimited by the segments $[AC]$ and $[CD]$, where the points $C$ and $D$ have coordinates

$$\text{C} \ : \ (h_2(p) \ , \ 1) \tag{42a}$$

$$\text{D} \ : \ (0 \ , \ 1 + 2h_2(p)) \ , \tag{42b}$$

and are obtained by evaluating the result of [18, Theorem 1] with the choices $U = S_0$ and $U = (S_0, S_1, S_2)$, respectively. The shaded surface represents the region of rate tuples which do not allow perfect secure transmission of the source component $S_0^n$.

## Appendix A
## Proof of Theorem 1

### A. Proof of Converse

Let $(R, R_1, D_1, \Delta)$ be an achievable rate-distortion-equivocation quadruple for the lossy Helper model with equivocation constraint of Figure 2a. Let $f$, $f_1$ and $f_2$ be the associated encoding functions at the encoder; and $g$ the decoding function at the legitimate receiver. That is,

$$W = f(S_0^n, S_1^n) \ , \tag{43}$$

$$W_1 = f_1(S_0^n, S_1^n) \ , \tag{44}$$

$$\hat{S}_1^n = g(W, W_1) \ , \tag{45}$$

with

$$\mathbb{E}d_1(S_1^n, \hat{S}_1^n) \leq nD_1 \ , \tag{46}$$

$$H(S_0^n, S_1^n|W) \geq n\Delta. \tag{47}$$

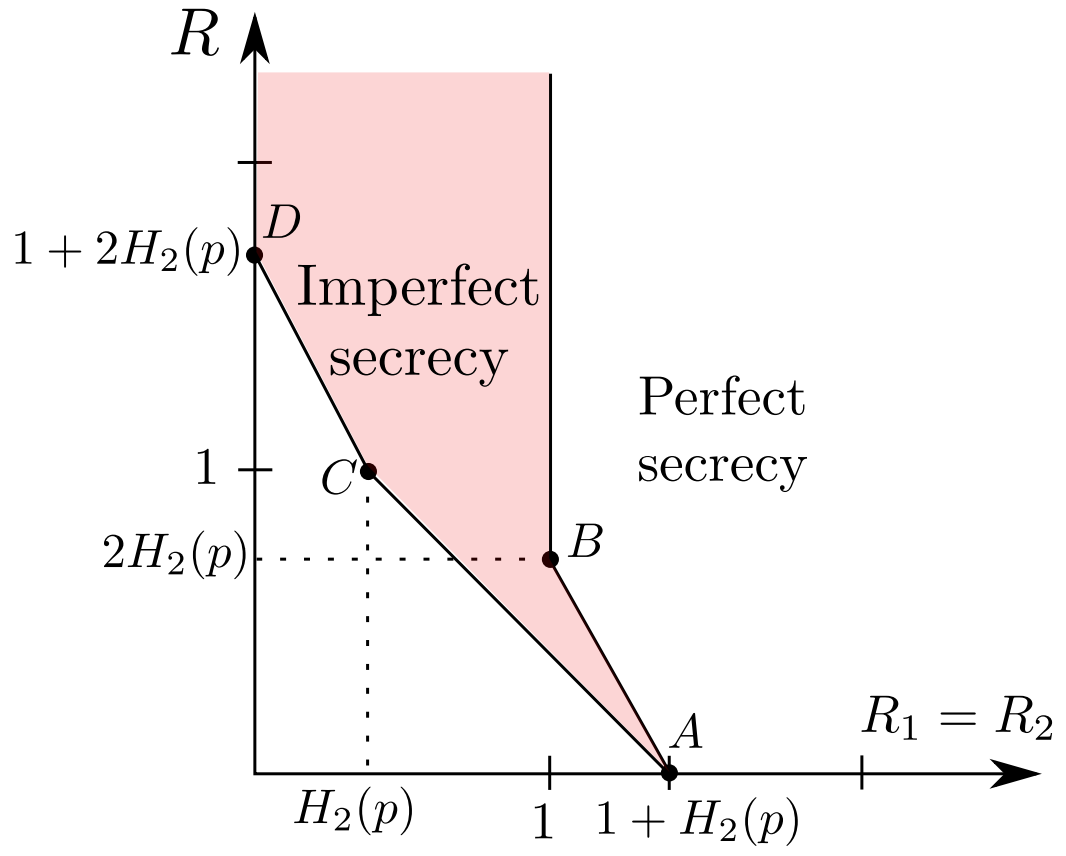

Fig. 9: Perfectly secure rate-region of the binary Gray-Wyner example of Figure 8.

Let, for $i = 1, \ldots, n$, the auxiliary random variable $U_i$ defined as

$$U_i \triangleq (W, S_0^{i-1}, S_1^{i-1}). \tag{48}$$

First, we bound the rate of message $W$ on the public link as follows,

$$
\begin{aligned}
n\,R &\geq H(W) && \text{(49a)}\\
&\geq I(W; S_0^n, S_1^n) && \text{(49b)}\\
&= \sum_{i=1}^{n} I(W; S_{0,i} S_{1,i} | S_0^{i-1} S_1^{i-1}) && \text{(49c)}\\
&\overset{(a)}{=} \sum_{i=1}^{n} I(W S_0^{i-1} S_1^{i-1}; S_{0,i} S_{1,i}) && \text{(49d)}\\
&\overset{(b)}{=} \sum_{i=1}^{n} I(U_i; S_{0,i} S_{1,i}) && \text{(49e)}
\end{aligned}
$$

where $(a)$ holds since the sources are memoryless and $(b)$ follows using (48).

Similarly, the rate of message $W_1$ on the private link can be bounded as

$$\begin{aligned}
n\,R_1 &\geq H(W_1) && \text{(50a)}\\
&\geq H(W_1|W) && \text{(50b)}\\
&\geq I(W_1; S_0^n S_1^n|W) && \text{(50c)}\\
&\overset{(a)}{=} I(W_1 \hat{S}_1^n; S_0^n S_1^n|W) && \text{(50d)}\\
&\geq I(\hat{S}_1^n; S_0^n S_1^n|W) && \text{(50e)}\\
&= \sum_{i=1}^{n} I(\hat{S}_1^n; S_{0,i} S_{1,i}|W S_0^{i-1} S_1^{i-1}) && \text{(50f)}\\
&\geq \sum_{i=1}^{n} I(\hat{S}_{1,i}; S_{0,i} S_{1,i}|W S_0^{i-1} S_1^{i-1}) && \text{(50g)}\\
&\overset{(b)}{=} \sum_{i=1}^{n} I(\hat{S}_{1,i}; S_{0,i} S_{1,i}|U_i) && \text{(50h)}
\end{aligned}$$

where $(a)$ holds since $\hat{S}_1^n = g(W, W_1)$ and $(b)$ follows using (48).
We now bound the the equivocation as,

$$\begin{aligned}
n\,\Delta &\leq H(S_0^n S_1^n|W) && \text{(51a)}\\
&= \sum_{i=1}^{n} H(S_{0,i} S_{1,i}|W S_0^{i-1} S_1^{i-1}) && \text{(51b)}\\
&= \sum_{i=1}^{n} H(S_{0,i} S_{1,i}|U_i) && \text{(51c)}
\end{aligned}$$

where the last equality follows by substituting using (48).
Let $Q$ be an integer-valued random variable, ranging from 1 to $n$, uniformly distributed over $[1:n]$ and independent of all other variables $(S_0, U, S_1)$. We have

$$\begin{aligned}
R &\geq \frac{1}{n} \sum_{i=1}^{n} I(U_i; S_{0,i} S_{1,i}) && \text{(52a)}\\
&= \sum_{i=1}^{n} \mathbb{P}(Q=i)\; I(U_i; S_{0,i} S_{1,i}|Q=i) && \text{(52b)}\\
&= I(U_Q; S_{0,Q} S_{1,Q}|Q) && \text{(52c)}\\
&\overset{(a)}{=} I(U_Q; S_0 S_1|Q) && \text{(52d)}\\
&\overset{(b)}{=} I(Q, U_Q; S_0 S_1) && \text{(52e)}
\end{aligned}$$

where $(a)$ holds since the sources are i.i.d, and $(b)$ holds since $Q$ is independent from all other variables.
Let us now define $U \triangleq (Q, U_Q)$. Letting $\hat{S}_1 = (Q, \hat{S}_{1,Q})$, we can easily show that

$$R_1 \geq I(\hat{S}_1; S_0 S_1|U) \quad (53)$$

$$\Delta \leq H(S_0 S_1|U)\;. \quad (54)$$

This completes the proof of converse of Theorem 1.

*B. Proof of Achievability*

We describe briefly the codebook generation, encoding and decoding operations. The analysis of the allowed equivocation will follow.

*Codebook Generation:*

i) Generate $2^{nR}$ independent $n$-sequences $u^n(w)$ indexed by $w = 1, \ldots, 2^{nR}$, each with i.i.d. elements drawn according to $\prod_{i=1}^{n} P_U(u_i(w))$.

ii) For each index $w$, generate $2^{nR_1}$ independent $n$-sequences $\hat{s}_1^n(w, w_1)$ indexed by $w_1 = 1, \ldots, 2^{nR_1}$, each with elements drawn i.i.d. according to $\prod_{i=1}^{n} P_{\hat{S}_1|U}(\hat{s}_{1,i}(w, w_1)|u_i(w))$.

*Encoding:* Upon observing the source pair $(s_0^n, s_1^n)$, the encoder finds a pair of sequences $\left(u^n(w), \hat{s}_1^n(w, w_1)\right)$ such that $u^n(w)$, $\hat{s}_1^n(w, w_1)$ and $(s_0^n, s_1^n)$ are jointly typical, i.e.,

$$\left(u^n(w), \hat{s}_1^n(w, w_1), s_0^n, s_1^n\right) \in \mathcal{T}^{(n)}_{[U\hat{S}_1 S_0 S_1]}. \tag{55}$$

If no such sequence $u^n(w)$ exists, set $w = 2^{nR} + 1$; and if no such sequence $\hat{s}_1^n(w, w_1)$ exists, set $w_1 = 2^{nR_1} + 1$. The encoder then transmits the message $w$ on the public link and the message $w_1$ on the private link.

*Estimation at Legitimate Receiver:* Upon receiving the pair of indices $(w, w_1)$, the decoder sets $\hat{s}_1^n(w, w_1)$ as the reconstructed version of the sequence $S_1^n$.

*Analysis of Equivocation:* Le $C$ denotes the used codebook, assumed to be known to the encoder, legitimate receiver and eavesdropper. We lower-bound the conditional entropy $H(S_0^n, S_1^n|W)$ as follows.

$$
\begin{aligned}
H(S_0^n, S_1^n|W) &= \sum_{w=1}^{1+2^{nR}} \mathbb{P}(W = w) H(S_0^n S_1^n | W = w) && \text{(56a)}\\
&\geq \sum_{w=1}^{2^{nR}} P_W(w) H(S_0^n S_1^n | W = w) && \text{(56b)}\\
&\overset{(a)}{=} \sum_{w=1}^{2^{nR}} P_W(w) H\left(S_0^n S_1^n | W = w, (S_0^n, S_1^n, U^n) \in \mathcal{T}^{(n)}_{[S_0 S_1 U]}\right) && \text{(56c)}\\
&\overset{(b)}{=} \sum_{w=1}^{2^{nR}} P_W(w) H\left(S_0^n S_1^n | U^n = u^n(w), (S_0^n, S_1^n, U^n) \in \mathcal{T}^{(n)}_{[S_0 S_1 U]}\right) && \text{(56d)}\\
&= \sum_{w=1}^{2^{nR}} P_W(w) H\left(S_0^n S_1^n | U^n = u^n(w), (S_0^n, S_1^n) \in \mathcal{T}^{(n)}_{[S_1 S_2 | u^n(w)]}\right) && \text{(56e)}\\
&= -\sum_{w=1}^{2^{nR}} P_W(w) \sum_{(s_0^n, s_1^n) \in \mathcal{T}^{(n)}_{[S_0 S_1 | u^n(w)]}} \mathbb{P}\left(S_0^n = s_0^n, S_1^n = s_1^n | U^n = u^n(w)\right)\\
&\qquad \times \log_2\left(\mathbb{P}\left(S_0^n = s_0^n, S_1^n = s_1^n | U^n = u^n(w)\right)\right) && \text{(56f)}\\
&\overset{(c)}{\geq} n \sum_{w=1}^{2^{nR}} P_W(w) \left[H(S_0 S_1 | U) - \epsilon_n\right] \sum_{(s_0^n, s_1^n) \in \mathcal{T}^{(n)}_{[S_0 S_1 | u^n(w)]}} \mathbb{P}\left(S_0^n = s_0^n, S_1^n = s_1^n | U^n = u^n(w)\right) && \text{(56g)}\\
&= n\left[H(S_0 S_1 | U) - \epsilon_n\right] \sum_{w=1}^{2^{nR}} P_W(w) && \text{(56h)}
\end{aligned}
$$

$$= n\ [H(S_0S_1|U) - \epsilon_n]\,(1 - P^n(e)) \tag{56i}$$

where $(a)$ follows by noticing that if $w \neq 2^{nR}+1$, then the sequences $u^n(w)$, $s_0^n$ and $s_1^n$ are jointly typical, i.e., $(s_0^n, s_1^n, u^n(w)) \in \mathcal{T}^{(n)}_{[S_0S_1U]}$; $(b)$ holds since by the code construction there is a one-to-one mapping between the set $\{u^n\}$ and the set of indices $\{w \in [1:2^{nR}]\}$; and $(c)$ holds by the conditional typicality lemma [32, Section 2.5].

## Appendix B
## Proof of Corollary 1

First, note that, in the case of lossless reconstruction of the source $S_1^n$ at the legitimate receiver, Theorem 1 reduces to the set of triples $(R, R_1, \Delta)$ that satisfy

$$R_1 \geq H(S_1|U) \tag{57a}$$

$$R \geq I(U; S_0, S_1) \tag{57b}$$

$$\Delta \leq H(S_0, S_1|U) \tag{57c}$$

for some conditional p.m.f. $P_{U|S_0,S_1}$. Let $\mathcal{R}(P_{U|S_0,S_1})$ denote such region, and

$$\mathcal{R} = \bigcup_{P_{U|S_0,S_1}} \mathcal{R}(P_{U|S_0,S_1}). \tag{58}$$

Also, let $\mathcal{R}'$ denote the region described by (12), i.e., $\mathcal{R}'$ is the set of triples $(R, R_1, \Delta)$ that satisfy

$$R + R_1 \geq H(S_1) \tag{59a}$$

$$\Delta \leq H(S_0|S_1) + \min\{\, R_1,\ H(S_1)\,\}. \tag{59b}$$

We want to show that

$$\mathcal{R} = \mathcal{R}'. \tag{60}$$

It is easy to see that $\mathcal{R} \subseteq \mathcal{R}'$. Let $(R, R_1, \Delta) \in \mathcal{R}'$, it remains to show that $(R, R_1, \Delta) \in \mathcal{R}$ for some choice of $P_{U|S_0,S_1}$. If $R_1 \geq H(S_1)$, then the triple $(R, R_1, \Delta)$ satisfies (57) with the choice $U = \emptyset$, i.e., $(R, R_1, \Delta) \in \mathcal{R}$. If $R_1 < H(S_1)$, there exists $\alpha \in [0, 1]$ such that $R_1 = \alpha H(S_1)$. Let $T$ be a Bernoulli-$\alpha$ random variable that is independent of $S_1$. Also, let $U = (T, U_T)$, where

$$U_{\{T=0\}} = S_1\ , \quad \text{and} \quad U_{\{T=1\}} = \emptyset. \tag{61}$$

It is easy to see that

$$H(S_1|U) = \alpha H(S_1) = R_1\ . \tag{62}$$

Similarly, since $R + R_1 \geq H(S_1)$, then we have

$$R \geq H(S_1) - R_1 \tag{63a}$$

$$= H(S_1) - H(S_1|U) \tag{63b}$$

$$= I(U; S_1) \tag{63c}$$

$$= I(U; S_0, S_1) \tag{63d}$$

where the last equality holds since the so-defined $U$ is such that $U -\!\!\!\circ\!\!\!- S_1 -\!\!\!\circ\!\!\!- S_0$ is a Markov chain. Also, since by assumption $\Delta \leq H(S_0|S_1) + \min\{R_1, H(S_1)\}$, we have

$$\Delta \leq H(S_0|S_1) + H(S_1|U) \tag{64a}$$

$$\stackrel{(a)}{=} H(S_0|U, S_1) + H(S_1|U) \tag{64b}$$

$$= H(S_0, S_1|U) \tag{64c}$$

where $(a)$ follows since $U$ satisfies the Markov chain $U -\!\!\!\circ\!\!\!- S_1 -\!\!\!\circ\!\!\!- S_0$.

The above means that the triple $(R, R_1, \Delta)$ satisfies (62), (63) and (64) for the given choice of $P_{U|S_0,S_1}$; and so $(R, R_1, \Delta) \in R$. This completes the proof of Corollary 1.

## Appendix C
## Proof of Theorem 2

### A. Proof of Converse

Let $(R, R_1, D_1, \Delta)$ be an achievable rate-distortion-equivocation quadruple for the Lossy Helper model of Figure 2b. Let $f$ and $f_1$ denote then the associated encoding functions and $g$ the decoding function at the legitimate receiver. That is,

$$W = f(S_0^n, S_1^n) \; , \tag{65}$$

$$W_1 = f_1(S_0^n, S_1^n) \; , \tag{66}$$

$$\hat{S}_1^n = g(W, W_1) \; , \tag{67}$$

where $\mathbb{E}d_1^{(n)}(S_1^n, \hat{S}_1^n) \leq D_1$.

Using Fano's inequality, the lossless reconstruction of the source $S_0^n$ at the legitimate receiver yields that there exists a sequence $\epsilon_n$, with $\lim_{n\to\infty} \epsilon_n = 0$, such that:

$$H(S_0^n|WW_1) \leq n\epsilon_n \; , \tag{68}$$

Besides, one has by definition that:

$$H(S_0^n|W) \geq n\Delta \; , \tag{69}$$

First, the sum-rate $(R + R_1)$ can be lower-bounded as

$$n\,(R + R_1) \geq H(W_1 W) \tag{70a}$$

$$\geq I(WW_1; S_0^n S_1^n) \tag{70b}$$

$$\geq I(WW_1; S_0^n) + I(WW_1; S_1^n|S_0^n) \tag{70c}$$

$$\stackrel{(a)}{\geq} I(WW_1; S_0^n) + I(WW_1\hat{S}_1^n; S_1^n|S_0^n) \tag{70d}$$

$$\geq I(WW_1; S_0^n) + I(\hat{S}_1^n; S_1^n|S_0^n) \tag{70e}$$

$$\stackrel{(b)}{\geq} nH(S_0) + I(\hat{S}_1^n; S_1^n|S_0^n) - n\,\epsilon_n \tag{70f}$$

$$= nH(S_0) + \sum_{i=1}^{n} I(\hat{S}_1^n; S_{1,i}|S_{0,i}S_1^{i-1}, S_0^{i-1}, S_{0,i+1}^n) - n\,\epsilon_n \tag{70g}$$

$$\stackrel{(c)}{=} nH(S_0) + \sum_{i=1}^{n} I(\hat{S}_1^n S_1^{i-1} S_0^{i-1} S_{0,i+1}^n ; S_{1,i}|S_{0,i}) - n\,\epsilon_n \tag{70h}$$

$$\geq nH(S_0) + \sum_{i=1}^{n} I(\hat{S}_{1,i}; S_{1,i}|S_{0,i}) - n\,\epsilon_n \ , \tag{70i}$$

where $(a)$ follows from that $\hat{S}_1^n = g(W, W_1)$, $(b)$ follows from (68), while $(c)$ is a consequence of the following Markov chain,

$$S_{1,i} -\!\!\circ\!\!- S_{0,i} -\!\!\circ\!\!- (S_1^{i-1}, S_0^{i-1}, S_{0,i+1}^n) \ . \tag{71}$$

which holds since the sources $(S_0^n, S_1^n)$ are memoryless.

Finally, the rate $R_1$ on the private link can be lower-bounded as

$$n\,R_1 \geq H(W_1) \geq I(S_0^n; W_1|W) \stackrel{(a)}{\geq} H(S_0^n|W) - n\,\epsilon_n \tag{72a}$$

$$\stackrel{(b)}{\geq} n\Delta - n\,\epsilon_n \ , \tag{72b}$$

where $(a)$ follows by using (68) and $(b)$ using (69).

The rest of the proof of converse follows by using standard single-letterization techniques, which we omit here for brevity. This completes the proof of the converse of Theorem 2.

## *B. Proof of Achievability*

The coding scheme that we use for the proof of the direct part of Theorem 2 uses a careful combination of binning and rate-splitting. Specifically, let $R_1 = R_{1,1} + R$. The codebook generation, encoding and decoding operations are as follows.

*Codebook Generation:*

i) Randomly and independently assign a pair of indices $(w_{0,0}, w_{1,0})$ to every $s_0^n \in \mathcal{T}_{[S_0]}^{(n)}$, where $(w_{0,0}, w_{1,0}) \in 2^{n(R_{0,0}+R_{1,0})}$. To every $s_0^n \notin \mathcal{T}_{[S_0]}^{(n)}$, assign $(w_{0,0}, w_{1,0}) = (2^{nR_{0,0}}+1, 2^{nR_{1,0}}+1)$.

ii) For each sequence $s_0^n \in \mathcal{T}_{[S_0]}^{(n)}$, generate $2^{n(R_{1,1}+R)}$ independent sequences $\hat{s}_1^n(w_{0,1}, w_{1,1})$, where $(w_{1,0}, w_{1,1}) \in 2^{n(R_{1,0}+R_{1,1})}$, each with i.i.d components drawn according to $\prod_{i=1}^{n} P_{\hat{S}_1|S_0}(\hat{s}_{1,i}(w_{0,1}, w_{1,1})|s_{0,i})$.

iii) Reveal the codebook to the legitimate receiver and eavesdropper.

*Encoding:* Upon observing $(s_0^n, s_1^n)$, the encoder first recovers $(w_{0,0}, w_{1,0})$ the bin indices of $s_0^n$. Then, it looks for a sequence $\hat{s}_1^n(w, w_1)$ jointly typical with the sources $(S_0^n, S_1^n)$, i.e

$$\left(s_0^n, s_1^n, \hat{s}_1^n\,(w_{0,1}, w_{1,1})\right) \in \mathcal{T}_{[S_0 S_1 \hat{S}_1]}^{(n)} \ . \tag{73}$$

If no such sequence exists, the encoder sets $w_{1,0} = 2^{nR_{1,0}} + 1$ and $w_{1,1} = 2^{nR_{1,1}} + 1$; if more than one exists, it selects at random of them. The encoder sends then the messages $w_1 = (w_{1,0}, w_{1,1})$ over the private link while the message $w = (w_{0,0}, w_{0,1})$ is sent through the public link.

*Decoding:* Based on all received indices, the decoder first recovers the sequence $S_0^n$ as the unique typical sequence in the bin indexed by $(w_{0,0}, w_{1,0})$. Then, knowing $s_0^n$, it sets its description as being $\hat{S}_1^n(w_{1,1}, w_{0,1})$.

The encoding and decoding are successful provided that $n$ is large enough and

$$R_{1,1} + R_{0,1} \geq I(S_1; \hat{S}_1|S_0) \ , \tag{74a}$$

$$R_{0,0} + R_{1,0} \geq H(S_0) \ . \tag{74b}$$

*Equivocation Analysis:* Let $(S_0^n, S_1^n)$ be the observed sequence and $W$ the message sent over the public link. For convenience let us define the index $\tilde{W} = (W_{0,1}, W_{1,1})$, of rate $\tilde{R} = R_{0,1}+R_{1,1}$. We start by writing that

$$H(S_0^n|W) = H(S_0^n|W_{0,0}, W_{0,1}) \geq H(S_0^n|W_{0,0}, W_{0,1}, W_{1,1}) \ . \tag{75}$$

The proof if two-fold and consists in establishing two main inequalities.

i) The first inqueality which writes as

$$H(S_0^n|\tilde{W}) \geq H(S_0) - 3n\epsilon_n \ , \tag{76}$$

amounts to stating that the index $\tilde{W}$ is independent of $S_0^n$.

ii) The second inequality writes as

$$H(S_0^n|W_{0,0}, \tilde{W}) \geq nR_{1,0} - n\epsilon_n \ , \tag{77}$$

and implies that the only secure part of $S_0^n$ is the index transmitted over the private link.

To prove the first inequality (76), we start by writing that

$$H(S_0^n|\tilde{W}) = H(S_0^n) + H(\tilde{W}|S_0^n) - H(\tilde{W}) \tag{78a}$$

$$\geq H(S_0^n) + H(\tilde{W}|S_0^n) - n\,\tilde{R} \ . \tag{78b}$$

The proof consists in proving that the difference term $H(\tilde{W}|S_0^n) - n\,\tilde{R}$ in the RHS of (78b) is arbitrary small. To this end, let us first denote $E$ the event that an encoding error occurs and denote by $\bar{E}$ its error complement. First, observe that

$$H(\tilde{W}|S_0^n) \geq \mathbb{P}(\bar{E})H(\tilde{W}|S_0^n, \bar{E}) \ , \tag{79}$$

which is a consequence of that if an error occurs, then $\tilde{W}$ can take only one possible value, i.e. $2^{n\tilde{R}}+1$.

The conditional entropy $H(\tilde{W}|S_0^n, \bar{E})$ on the RHS of (79) is given by

$$H(\tilde{W}|S_0^n, \bar{E}) = -\sum_{s_0^n \in \mathcal{T}_{[S_0]}^{(n)}} \sum_{\tilde{w}=1}^{2^{n\tilde{R}}} \mathbb{P}(\tilde{W} = \tilde{w}, S_0^n = s_0^n|\bar{E}) \log_2\left(\mathbb{P}(\tilde{W} = \tilde{w}|S_0^n = s_0^n, \bar{E})\right), \tag{80}$$

with, for given $s_0^n$ and $\tilde{w} \in [1 : 2^{n\tilde{R}}]$,

$$\mathbb{P}(\tilde{W} = \tilde{w}|S_0^n = s_0^n, \bar{E}) = \sum_{s_1^n \in \mathcal{T}_{[S_1|s_0^n]}^{(n)}} P_{S_1|S_0}^n(s_1^n|s_0^n)\mathbb{P}(\tilde{W} = \tilde{w}|S_1^n = s_1^n, S_0^n = s_0^n, \bar{E}). \tag{81}$$

In what follows, we compute the conditional probability $\mathbb{P}(\tilde{W} = \tilde{w}|S_1^n = s_1^n, S_0^n = s_0^n, \bar{E})$ for given $(s_0^n, s_1^n)$. Let, for a given $s_0^n$, $\hat{\mathcal{S}}_1^n(s_0^n)$ denote the set of generated sequences $\{\hat{s}_1^n(\tilde{w})\}$ of the codebook, among all possible sequences in $\mathcal{T}_{[\hat{S}_1|s_0^n]}^{(n)}$. For a given pair $(s_0^n, s_1^n)$, if no error occurs during the encoding step, the encoder chooses *at random* one sequence $\hat{s}_1^n(\tilde{w})$ among all those that are strongly jointly typical with the pair $(s_0^n, s_1^n)$. That is, the sequence $\hat{s}_1^n(\tilde{w})$ is chosen at random in the following intersection set

$$\mathcal{T}_{[\hat{S}_1|s_1^n, s_0^n]}^{(n)} \cap \hat{\mathcal{S}}_1^n(s_0^n) \ . \tag{82}$$

For given $(s_0^n, s_1^n)$, every sequence from this intersection set has probability

$$\mathbb{P}(\tilde{W} = \tilde{w}|S_1^n = s_1^n, S_0^n = s_0^n, \bar{E}) = \|\mathcal{T}^{(n)}_{[\hat{S}_1|s_1^n,s_0^n]} \cap \hat{\mathcal{S}}_1^n(s_0^n)\|^{-1} \times \mathbb{1}\left(\left(\hat{s}_1^n(\tilde{w}), s_0^n, s_1^n\right) \in \mathcal{T}^{(n)}_{[\hat{S}_1,S_0,S_1]}\right). \tag{83}$$

Key to the rest of the proof is a counting argument that we use to bound the cardinality of the set $\mathcal{T}^{(n)}_{[\hat{S}_1|s_1^n,s_0^n]} \cap \hat{\mathcal{S}}_1^n(s_0^n)$. The result is stated in the following lemma, whose proof is relegated to Appendix D. At this stage, we mention that a different approach to computing a probability that is similar to (83) can be found in [35], and was also used later on in [36, Appendix A].

**Lemma 1** (Encoding set cardinality bound). *If $\tilde{R} \geq I(\hat{S}_1; S_1|S_0) + 2\epsilon_n$, the cardinality of the encoding set $\mathcal{T}^{(n)}_{[\hat{S}_1|s_1^n,s_0^n]} \cap \hat{\mathcal{S}}_1^n(s_0^n)$ satisfies*

$$(1-\epsilon)2^{n[\tilde{R}-I(\hat{S}_1;S_1|S_0)-2\epsilon_n]} \leq \left\|\mathcal{T}^{(n)}_{[\hat{S}_1|s_1^n,s_0^n]} \cap \hat{\mathcal{S}}_1^n(s_0^n)\right\| \leq (1+\epsilon)2^{n[\tilde{R}-I(\hat{S}_1;S_1|S_0)-2\epsilon_n]} . \tag{84}$$

*Proof.* The proof of Lemma 1 is given in Appendix D. □

We continue with the analysis of the equivocation. Using (84), equation (83) gives

$$\mathbb{P}(\tilde{W} = \tilde{w}|S_1^n = s_1^n, S_0^n = s_0^n, \bar{E}) \leq 2^{n[-\tilde{R}+I(\hat{S}_1;S_1|S_0)+2\epsilon_n]} \times \mathbb{1}\left((\hat{s}_1^n(\tilde{w}), s_0^n, s_1^n) \in \mathcal{T}^{(n)}_{[\hat{S}_1,S_0,S_1]}\right) . \tag{85}$$

Thus, the conditional probability of (81) satisfies

$$\mathbb{P}(\tilde{W} = \tilde{w}|S_0^n = s_0^n, \bar{E}) = \sum_{s_1^n \in \mathcal{T}^{(n)}_{[S_1|s_0^n]}} P^n_{S_1|S_0}(s_1^n|s_0^n)\ \mathbb{P}(\tilde{W} = \tilde{w}|S_1^n = s_1^n, S_0^n = s_0^n, \bar{E}) \tag{86a}$$

$$\leq \sum_{s_1^n \in \mathcal{T}^{(n)}_{[S_1|\hat{s}_1^n(\tilde{w}),s_0^n]}} P^n_{S_1|S_0}(s_1^n|s_0^n)\ 2^{n[-\tilde{R}+I(\hat{S}_1;S_1|S_0)+2\epsilon_n]} \tag{86b}$$

$$\leq 2^{n\left[H(S_1|S_0\hat{S}_1)-H(S_1|S_0)-\tilde{R}+I(\hat{S}_1;S_1|S_0)+3\epsilon_n\right]} \tag{86c}$$

$$\leq 2^{n[-\tilde{R}+3\epsilon_n]}, \tag{86d}$$

where (86c) follows by using (85).

Continuing from (80), we then get

$$H(\tilde{W}|S_0^n, \bar{E}) = -\sum_{s_0^n \in \mathcal{T}^{(n)}_{[S_0]}} \sum_{\tilde{w}=1}^{2^{n\tilde{R}}} \mathbb{P}(\tilde{W} = \tilde{w}, S_0^n = s_0^n|\bar{E})\ \log_2\left(\mathbb{P}(\tilde{W} = \tilde{w}|S_0^n = s_0^n, \bar{E})\right) \tag{87a}$$

$$\geq n \sum_{\tilde{w}=1}^{2^{nR}} \sum_{s_0^n \in \mathcal{T}^{(n)}_{[S_0]}} \mathbb{P}(\tilde{W} = \tilde{w}, S_0^n = s_0^n|\bar{E})\ \left(\tilde{R} - 3\epsilon_n\right) \tag{87b}$$

$$\geq n[\tilde{R} - 3\epsilon_n], \tag{87c}$$

where (87b) follows by using (86d).

Finally, using (79) and (87c), we get that the conditional entropy $H(S_0^n|\tilde{W})$ satisfies

$$H(S_0^n|\tilde{W}) \geq nH(S_0) + H(\tilde{W}|S_0^n, \bar{E}) - n\,\tilde{R} \tag{88}$$

$$\geq nH(S_0) - 3n\epsilon_n. \tag{89}$$

Now that we have proved that the index $\tilde{W}$ is independent of $S_0^n$, we move to prove the second inequality (77). To this end, observe that:

$$
\begin{aligned}
H(S_0^n|W_{0,0},\tilde{W}) &= H(S_0^n|W_{0,0}) - I(S_0^n;\tilde{W}|W_{0,0}) && \text{(90a)}\\
&\geq H(S_0^n|W_{0,0}) - I(S_0^n W_{0,0};\tilde{W}) && \text{(90b)}\\
&\overset{(a)}{=} H(S_0^n|W_{0,0}) - I(S_0^n;\tilde{W}) && \text{(90c)}\\
&\overset{(b)}{\geq} H(S_0^n|W_{0,0}) - 3n\epsilon_n && \text{(90d)}\\
&= H(S_0^n W_{1,0}|W_{0,0}) - H(W_{1,0}|S_0^n W_{0,0}) - 3n\epsilon_n && \text{(90e)}\\
&\overset{(c)}{\geq} H(S_0^n W_{1,0}|W_{0,0}) - 3n\epsilon_n && \text{(90f)}\\
&\geq H(W_{1,0}|W_{0,0}) - 3n\epsilon_n && \text{(90g)}\\
&\overset{(d)}{=} H(W_{1,0}) - 3n\epsilon_n && \text{(90h)}\\
&= nR_{1,0} - 3n\epsilon_n \ , && \text{(90i)}
\end{aligned}
$$

where $(a)$ is a consequence of that, knowing the codebook, $W_{0,0}$ is a function of $S_0^n$, while $(b)$ stems from (89). As for $(c)$, it results from the fact that $W_{1,0}$ is a function of $S_0^n$, and $(d)$ is a consequence of the random binning procedure.

Last, note a trivial bound on the leakage is given by $H(S_0)$ since $1/nH(S_0^n|W) \leq H(S_0)$, thus the bound on equivocation can be rewritten as:

$$H(S_0^n|W_{0,0},\tilde{W}) \geq n\min\{R_{1,0},H(S_0)\} - 3n\epsilon_n \ , \tag{91}$$

*Fourier Motzkin Elimination:* Summarizing, the triple $(R,R_1,D_1,\Delta)$ is achievable if there exists $(R_{0,0},R_{0,1},R_{1,0},R_{1,1})$ such that

$$
\begin{aligned}
R_{1,1} + R_{0,1} &\geq I(S_1;\hat{S}_1|S_0) \ , && \text{(92a)}\\
R_{0,0} + R_{1,0} &\geq H(S_0) \ , && \text{(92b)}\\
\Delta &\leq \min\{R_{1,0},H(S_0)\} && \text{(92c)}\\
R_1 &= R_{1,1} + R_{1,0} && \text{(92d)}\\
R &= R_{0,0} + R_{0,1} && \text{(92e)}\\
0 &\leq \{R_{1,0},R_{1,1}\} \leq R_1 && \text{(92f)}\\
0 &\leq \{R_{0,0},R_{0,1}\} \leq R && \text{(92g)}
\end{aligned}
$$

Using Fourier-Motzkin elimination to successively project out $R_{0,0}$, $R_{0,1}$, $R_{1,0}$ and $R_{1,1}$), we get the region of Theorem 2. This completes the proof of Theorem 2.

## Appendix D
## Proof of Lemma 1

Recall the definition of the set $\mathcal{T}^{(n)}_{[\hat{S}_1|s_1^n,s_0^n]}\cap\hat{\mathcal{S}}_1^n(s_0^n)$ as given in Appendix C. (See Figure 10 for a schematic representation). The cardinality of this set is a random variable, which we denote hereafter as $C$, i.e.,

$$C \triangleq \|\mathcal{T}^{(n)}_{[\hat{S}_1|s_1^n,s_0^n]} \cap \hat{\mathcal{S}}_1^n(s_0^n)\| \ . \tag{93}$$

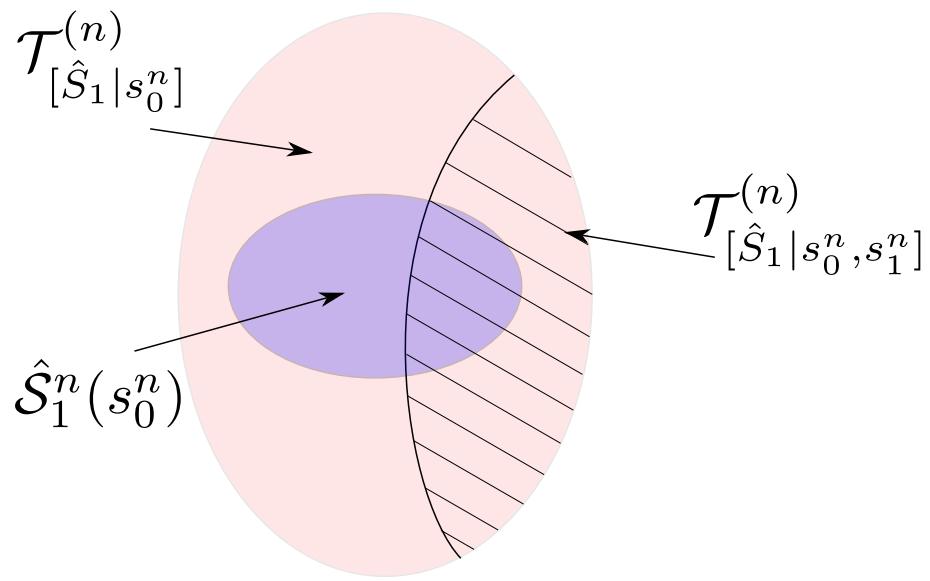

Fig. 10: Encoding set for a fixed pair $(s_0^n, s_1^n)$

Using Tchebychev-Bienayme's inequality, one gets that, for $\epsilon \geq 0$,

$$\mathbb{P}\Big( |C - \mathbb{E}(C)| \geq \epsilon \mathbb{E}(C) \Big) \leq \frac{\text{Var}(C)}{\epsilon^2 \mathbb{E}^2(C)} \ , \tag{94}$$

where Var($C$) designates the variance of the random variable $C$.

For convenience, let the following substitutions and notations,

$$N \triangleq \|\mathcal{T}^{(n)}_{[\hat{S}_1|s_0^n]}\| \ , \ B \triangleq \|\mathcal{T}^{(n)}_{[\hat{S}_1|s_1^n,s_0^n]}\| \ , \quad \text{and} \quad K \triangleq \|\hat{\mathcal{S}}_1^n(s_0^n)\| = 2^{n\tilde{R}} \ . \tag{95}$$

The random variable $C$ follows a *Hyper-Geometric* distribution; and the probability that it takes some value $j$, with $1 \leq j \leq \min\{B, K\}$, is given by

$$P_C(j) = \begin{pmatrix} B \\ j \end{pmatrix} \begin{pmatrix} N - B \\ K - j \end{pmatrix} \begin{pmatrix} N \\ K \end{pmatrix}^{-1} \ . \tag{96}$$

The expectation and variance of this variable are given respectively by

$$\mathbb{E}(C) = \frac{BK}{N} \ , \quad \text{and} \quad \text{Var}(C) = \frac{BK}{N} \frac{N - B}{N} \frac{N - K}{N - 1} \ . \tag{97}$$

Substituting in (97) using (95), and using the joint typicality lemma [32, page 29] and the conditional typicality lemma [32, page 27], we get

$$2^{n[\tilde{R} - I(\hat{S}_1;S_1|S_0) - 2\epsilon_n]} \leq \mathbb{E}(C) \leq 2^{n[\tilde{R} - I(\hat{S}_1;S_1|S_0) + 2\epsilon_n]} \ , \tag{98a}$$

$$\text{Var}(C) \leq \frac{BK}{N} \leq 2^{n[\tilde{R} - I(\hat{S}_1;S_1|S_0) + 2\epsilon_n]} \ . \tag{98b}$$

Next, using (98), the inequality (94) gives that, for all $\epsilon \geq 0$, we have

$$\mathbb{P}\Big(|C - \mathbb{E}(C)| \geq \epsilon \mathbb{E}(C)\Big) \leq \frac{1}{\epsilon^2} \ 2^{-n[\tilde{R} - I(\hat{S}_1;S_1|S_0) - 2\epsilon_n]} \ . \tag{99}$$

The above means that, if $\tilde{R} \geq I(\hat{S}_1; S_1|S_0) + 2\epsilon_n$, then for $\epsilon \geq 0$, the probability $\mathbb{P}\Big(|C - \mathbb{E}(C)| < \epsilon \mathbb{E}(C)\Big)$ tends to unity as $n \longrightarrow \infty$. Finally using (98), this leads to the desired result; and completes the proof of Lemma 1. □

## Appendix E

## Encryption-based coding scheme

In this section we show that, if the optimal rate-distortion region given in Theorem 2 suggests that the source $S_0$ be compressed over the private link of rate $R_1$, one could equally optimally transmit the source $S_0$ on the common link, provided a key of rate at least equal to $H(S_0)$ be used to encrypt it. Thus, the communication on the public link is not prevented, however, it has to be secured.

We describe hereafter the corresponding encryption-based coding scheme in which the private link is used as a private key to encrypt the common link transmission. The codebook generation, encoding and decoding are as follows.

*Codebook generation:*

i) Randomly and independently assign to each $s_0^n \in \mathcal{T}_{[S_0]}^{(n)}$ a triplet of bin indices $(\tilde{w}_0, w_{0,0}, w_{1,0}) \in [1 : 2^{n\tilde{R}_0}] \times [1 : 2^{nR_{0,0}}] \times [1 : 2^{nR_{1,0}}]$. Assign to each $s_0^n \notin \mathcal{T}_{[S_0]}^{(n)}$ an error triplet given by $(\tilde{w}_0, w_{0,0}, w_{1,0}) = (2^{n\tilde{R}_0} + 1, 2^{nR_{0,0}} + 1, 2^{nR_{1,0}} + 1)$.

ii) For each $s_0^n$, generate independently $2^{n(\tilde{R}_1+R1,1+R_{0,1})}$ sequences $\hat{s}_1^n(\tilde{w}_1, w_{1,1}, w_{0,1})$ where $\tilde{w}_1, w_{1,1}, w_{0,1} \in [1 : 2^{n\tilde{R}_1}]\times[1 : 2^{nR_{1,1}}]\times[1 : 2^{nR_{0,1}}]$, with i.i.d components drawn according to $\prod_{i=1}^{n} P_{\hat{S}_1|S_0}(\hat{s}_{1,i}(\tilde{w}_1, w_{1,1}, w_{0,1})|s_{0,i})$.

*Encoding:* Upon observing $(s_0^n, s_1^n)$, the encoder recovers the indices assigned to $s_0^n$, i.e. $(\tilde{w}_0, w_{0,0}, w_{1,0})$, and then looks for a triplet of indices $(\tilde{w}_1, w_{1,1}, w_{0,1})$ such that:

$$\left(\hat{s}_1^n(\tilde{w}_1, w_{1,1}, w_{0,1}), s_0^n, s_1^n\right) \in \mathcal{T}_{[\hat{S}_1 S_0 S_1]}^{(n)} , \tag{100}$$

It then transmits the indices $(\tilde{w}_1, w_{1,1}, w_{1,0})$ on the private link and transmits the indices $(\tilde{w}_0 \oplus \tilde{w}_1, w_{0,0}, w_{0,1})$ on the public link, where the XOR operation $\oplus$ is performed bit-wise to encrypt the index $\tilde{w}_0$.

*Decoding:* Based on the received indices, the decoder first recovers the sequence $s_0^n$ and then, recovers $s_1^n$ similarly.

The decoding is successful if:

$$\tilde{R}_0 + R_{0,0} + R_{1,0} \geq H(S_0) \tag{101a}$$

$$\tilde{R}_1 + R_{0,1} + R_{1,1} \geq I(\hat{S}_1; S_1|S_0) \tag{101b}$$

Besides, in order for the encryption to be successful, we impose that:

$$\tilde{R}_1 \geq \tilde{R}_0 . \tag{102}$$

In the following, we analyse the resulting equivocation, or equivalently, the induced rate leakage, at the eavesdropper about the source $s_0^n$ to be secured,

*Equivocation Analysis:* The leakage at the eavesdropper could result from the three indices transmitted on the public link, i.e. $\tilde{W}_0 \oplus \tilde{W}_1$, $W_{0,1}$ and $W_{0,0}$.

In the following, we bound the leakage of each of the aforementioned indices.

We start the analysis by writing:

$$I(W; S_0^n) = I\left(\tilde{W}_0 \oplus \tilde{W}_1, W_{0,0}, W_{0,1}; S_0^n\right) \tag{103a}$$

$$= I\left(\tilde{W}_0 \oplus \tilde{W}_1; S_0^n\right) + I\left(W_{0,1}; S_0^n|\tilde{W}_0 \oplus \tilde{W}_1\right) + I\left(W_{0,0}; S_0^n|W_{0,1}, \tilde{W}_0 \oplus \tilde{W}_1\right) . \tag{103b}$$

To bound the three RHS terms of (103b), we will introduce four crucial results.

- The first result we will resort to consists in stating that, similarly to the equivocation analysis in the proof of Theorem 2 in Appendix C, $(\tilde{W}_1, W_{0,1}, W_{1,1})$ are asymptotically independent of $s_0^n$, i.e. there exists $\epsilon_n$ such that:

$$I\left(\tilde{W}_1 W_{0,1} W_{1,1}; S_0^n\right) \leq n\epsilon_n . \tag{104}$$

- Next, since $\tilde{W}_1$ is independent of $S_0^n$, and since $\tilde{W}_0$ is a function of $S_0^n$, then, following Shannon's one time-pad proof, there exists $\epsilon_n$ such that:

$$I(\tilde{W}_0 \oplus \tilde{W}_1; \tilde{W}_0) \leq n\epsilon_n \tag{105}$$

- Since $\hat{s}_1^n$ are chosen randomly in the set of conditionally typical sequences $\mathcal{T}^{(n)}_{[\hat{S}_1|s_1^n, s_0^n]}$, then the two sub-indices appear as independent of each other

$$I\left(W_{0,1}; \tilde{W}_1|S_0^n\right) \leq n\epsilon_n \ . \tag{106}$$

- Finally, since $s_0^n$ are assigned at random to their bin indices, then for all $(\tilde{w}_0, w_{0,0}, w_{1,0}) \in \left[1 : 2^{n\tilde{R}_0}\right] \times \left[1 : 2^{nR_{0,0}}\right] \times \left[1 : 2^{nR_{1,0}}\right]$,

$$\mathbb{P}(\tilde{W}_0 = \tilde{w}_0, W_{0,0} = w_{0,0}, W_{1,0} = w_{1,0}) = 2^{-n\tilde{R}_0} 2^{-nR_{0,0}} 2^{-nR_{1,0}} \ , \tag{107}$$

  which implies that:

$$H(\tilde{W}_0, W_{0,1}|W_{0,0}) = H(\tilde{W}_0) + H(W_{0,1}) \ . \tag{108}$$

We proceed with the analysis as follows:

$$\begin{aligned}
I(\tilde{W}_0 \oplus \tilde{W}_1; S_0^n) &\leq I(\tilde{W}_0 \oplus \tilde{W}_1; S_0^n \tilde{W}_0) && \text{(109a)}\\
&= I(\tilde{W}_0 \oplus \tilde{W}_1; S_0^n|\tilde{W}_0) + I(\tilde{W}_0 \oplus \tilde{W}_1; \tilde{W}_0) && \text{(109b)}\\
&\overset{(a)}{\leq} I(\tilde{W}_0 \oplus \tilde{W}_1; S_0^n|\tilde{W}_0) + n\epsilon_n && \text{(109c)}\\
&= I(\tilde{W}_1; S_0^n|\tilde{W}_0) && \text{(109d)}\\
&\leq I(\tilde{W}_1; S_0^n) && \text{(109e)}\\
&\overset{(b)}{\leq} n\epsilon_n && \text{(109f)}
\end{aligned}$$

where $(a)$ is a result of (105) while $(b)$ results from (104).

As for the second term of the RHS of (103b), note that:

$$\begin{aligned}
I\left(W_{0,1}; S_0^n|\tilde{W}_0 \oplus \tilde{W}_1\right) &\leq I\left(W_{0,1}; S_0^n, \tilde{W}_0 \oplus \tilde{W}_1\right) && \text{(110a)}\\
&= I\left(W_{0,1}; S_0^n\right) + I\left(W_{0,1}; \tilde{W}_0 \oplus \tilde{W}_1|S_0^n\right) && \text{(110b)}\\
&\overset{(a)}{=} n\epsilon_n + I\left(W_{0,1}; \tilde{W}_1|S_0^n\right) && \text{(110c)}\\
&\overset{(b)}{\leq} 2n\epsilon_n \ . && \text{(110d)}
\end{aligned}$$

where $(a)$ results from (104) and $(b)$ stems from (106).

Now, to bound the last term of the RHS of (103b), we can write that:

$$I\left(W_{0,0};S_0^n|W_{0,1},\tilde{W}_0\oplus\tilde{W}_1\right)=H\left(S_0^n|W_{0,1},\tilde{W}_0\oplus\tilde{W}_1\right)-H\left(S_0^n|W_{0,0},W_{0,1},\tilde{W}_0\oplus\tilde{W}_1\right) \tag{111a}$$
$$\leq H(S_0^n)-H\left(S_0^n|W_{0,0},W_{0,1},\tilde{W}_0\oplus\tilde{W}_1\right) \tag{111b}$$
$$\overset{(a)}{=} H(S_0^n)-H(S_0^n|W_{0,0})+n\epsilon_n \tag{111c}$$
$$= H(S_0^n)-H(S_0^nW_{1,0}\tilde{W}_0|W_{0,0})+H(W_{1,0}\tilde{W}_0|S_0^nW_{0,0})+n\epsilon_n \tag{111d}$$
$$\overset{(b)}{=} H(S_0^n)-H(S_0^nW_{1,0}\tilde{W}_0|W_{0,0})+n\epsilon_n \tag{111e}$$
$$\leq H(S_0^n)-H(W_{1,0}\tilde{W}_0|W_{0,0})+n\epsilon_n \tag{111f}$$
$$\overset{(c)}{=} nH(S_0)-n(R_{1,0}+\tilde{R}_0)+n\epsilon_n \tag{111g}$$

where $(a)$ follows from that $(W_{0,1},\tilde{W}_0\oplus\tilde{W}_1)$ are almost independent of $(S_0^n,W_{0,0})$, and $(b)$ is a result of that the pair of indices $(W_{1,0},\tilde{W}_0)$ are a function of $S_0^n$, while $(c)$ is a consequence of (108).

Combining thus the inequalities in (109f), (110d) and (111g), we can write:

$$I(W;S_0^n)\leq nH(S_0)-n(R_{1,0}+\tilde{R}_0)+4n\epsilon_n \tag{112}$$

which implies that:

$$H(S_0^n|W)\geq n(R_{1,0}+\tilde{R}_0)-4n\epsilon_n \tag{113}$$

which ends the equivocation analysis.

*Fourier-Motzkin Elimination:* We resort to Fourier-Motzkin Elimination on the set of inequalities and equalities given by:

$$R_1=\tilde{R}_1+R_{1,0}+R_{1,1} \tag{114a}$$
$$R=\tilde{R}_0+R_{0,0}+R_{0,1} \tag{114b}$$
$$\tilde{R}_0+R_{0,0}+R_{1,0}\geq H(S_0) \tag{114c}$$
$$\tilde{R}_1+R_{0,1}+R_{1,1}\geq I(\hat{S}_1;S_1|S_0) \tag{114d}$$
$$\tilde{R}_1\geq\tilde{R}_0 \tag{114e}$$
$$\Delta\leq R_{1,0}+\tilde{R}_0\ , \tag{114f}$$

along with positivity constraints, to obtain the desired rate region:

$$R+R_1\geq H(S_0)+I(\hat{S}_1;S_1|S_0) \tag{115a}$$
$$R+R_1\geq\Delta+I(\hat{S}_1;S_1|S_0) \tag{115b}$$
$$\Delta\leq\min\{H(S_0),R_1\} \tag{115c}$$

# Appendix F
# Proof of Theorem 4

## A. Proof of Converse

Let $(R,R_1,D_1,D_2,\Delta)$ be an achievable rate-distortion-equivocation tuple for the lossy Gray-Wyner model with equivocation constraint of Figure 3a. Let $f$, $f_1$ and $f_2$ be the associated encoding functions at the encoder; and $g_1$ and $g_2$ the decoding functions at the legitimate receivers. That is,

$$W = f(S_1^n, S_2^n) \tag{116}$$

$$W_1 = f_1(S_1^n, S_2^n) \tag{117}$$

$$W_2 = f_2(S_1^n, S_2^n) \tag{118}$$

$$\hat{S}_1^n = g_1(W, W_1) \tag{119}$$

$$\hat{S}_2^n = g_2(W, W_2), \tag{120}$$

with

$$\mathbb{E}d_1^{(n)}(S_1^n, \hat{S}_1^n) \leq nD_1, \quad \mathbb{E}d_2^{(n)}(S_2^n, \hat{S}_2^n) \leq nD_2 \quad \text{and} \quad H(S_1^n, S_2^n|W) \geq n\Delta. \tag{121}$$

Let, for $i = 1, \ldots, n$, the auxiliary random variable $U_i$ defined as

$$U_i \triangleq (W, S_1^{i-1}, S_2^{i-1}). \tag{122}$$

First, we lower bound the rate $R$ of message $W$ as follows,

$$n\,R \geq H(W) \tag{123a}$$

$$\geq I(W; S_1^n S_2^n) \tag{123b}$$

$$= \sum_{i=1}^{n} I(W; S_{1,i} S_{2,i} | S_1^{i-1} S_2^{i-1}) \tag{123c}$$

$$= \sum_{i=1}^{n} I(W S_1^{i-1} S_2^{i-1}; S_{1,i} S_{2,i}) \tag{123d}$$

$$= \sum_{i=1}^{n} I(U_i; S_{1,i} S_{2,i}) \tag{123e}$$

where the last equality follows by substituting using (122).

Similarly, for $j = 1, 2$, we lower bound the rate $R_j$ of message $W_j$ as

$$n\,R_j \geq H(W_j) \tag{124a}$$

$$\geq H(W_j|W) \tag{124b}$$

$$\geq I(W_j; S_1^n S_2^n | W) \tag{124c}$$

$$\overset{(a)}{=} I(W_j \hat{S}_j^n; S_1^n S_2^n | W) \tag{124d}$$

$$\geq I(\hat{S}_j^n; S_1^n S_2^n | W) \tag{124e}$$

$$= \sum_{i=1}^{n} I(\hat{S}_j^n; S_{1,i} S_{2,i} | W S_1^{i-1} S_2^{i-1}) \tag{124f}$$

$$\geq \sum_{i=1}^{n} I(\hat{S}_{j,i}; S_{1,i} S_{2,i} | W S_1^{i-1} S_2^{i-1}) \tag{124g}$$

$$\overset{(b)}{=} \sum_{i=1}^{n} I(\hat{S}_{j,i}; S_{1,i} S_{2,i} | U_i)\ , \tag{124h}$$

where $(a)$ holds since $\hat{S}_j^n = g_j(W, W_j)$; and $(b)$ follows by substituting using (122).

The equivocation can be upper-bounded easily, as

$$n\,\Delta \le H(S_1^n S_2^n|W) \tag{125a}$$

$$= \sum_{i=1}^{n} H(S_{1,i}S_{2,i}|WS_1^{i-1}S_2^{i-1}) \tag{125b}$$

$$= \sum_{i=1}^{n} H(S_{1,i}S_{2,i}|U_i), \tag{125c}$$

where, similarly to the above, the last equality follows by substituting using (122).
The rest of the proof of converse follows using standard single-letterization techniques, and is omitted for brevity. □

*B. Proof of Achievability*

The coding scheme that we use for the proof of the direct part of Theorem 4 is a straightforward extension of that of the Helper model with equivocation on both sources in Appendix A to the case of two legitimate receivers. For this reason, hereafter, we only outline the main steps, and omit the details.

*Codebook Generation:*

i) Generate $2^{nR}$ independent $n$-sequences $u^n(w)$ indexed by $w = 1, \dots, 2^{nR}$, each with i.i.d elements drawn according to $\prod_{i=1}^{n} P_U(u_i(w))$.
ii) For each $w \in [1:2^{nR}]$, generate $2^{nR_1}$ independent $n$-sequences $\hat{s}_1^n(w, w_1)$ indexed by $w_1 = 1, \dots, 2^{nR_1}$, each with i.i.d. elements drawn according to the conditional $\prod_{i=1}^{n} P_{\hat{S}_1|U}(\hat{s}_{1,i}(w, w_1)|u_i(w))$.
iii) Similarly, for each $w \in [1:2^{nR}]$, generate $2^{nR_2}$ independent $n$-sequences $\hat{s}_2^n(w, w_2)$ indexed by $w_2 = 1, \dots, 2^{nR_2}$, each with i.i.d. elements drawn according to the conditional $\prod_{i=1}^{n} P_{\hat{S}_2|U}(\hat{s}_{2,i}(w, w_2)|u_i(w))$.

*Encoding:* Upon observing a pair of sources $(s_1^n, s_2^n)$, the encoder first finds an index $w \in [1:2^{nR}]$ such that $u^n(w)$ is jointly typical with the pair $(s_1^n, s_2^n)$, i.e.,

$$\left(u^n(w), s_1^n, s_2^n\right) \in \mathcal{T}^{(n)}_{[US_1S_2]} \;. \tag{126}$$

The encoding at this step can be performed with vanishing probability of error as long as $n$ is large and

$$R \ge I(U; S_1S_2). \tag{127}$$

Then, the encoder looks for a pair of sequences $\hat{s}_1^n(w, w_1)$ and $\hat{s}_2^n(w, w_2)$ such that, for $j = 1, 2$, $\hat{s}_j^n(w, w_j)$ is jointly typical with the triple $(u^n(w), s_1^n, s_2^n)$, i.e.,

$$\left(u^n(w), s_1^n, s_2^n, \hat{s}_j^n(w, w_j)\right) \in \mathcal{T}^{(n)}_{[S_1S_2U\hat{S}_j]} \;. \tag{128}$$

Similarly, the encoding at this step can be performed successfully as long as $n$ is large and

$$R_1 \ge I(\hat{S}_1; S_1S_2|U) \tag{129a}$$

$$R_2 \ge I(\hat{S}_2; S_1S_2|U) \;. \tag{129b}$$

The encoder transmits the index $w$ on the common public link, the index $w_1$ on the private link to the first legitimate receiver and the index $w_2$ on the private link to the second legitimate receiver.

*Estimation:* Both legitimate receivers get the index $w$ on the common public link; and Legitimate Receiver $j$, $j = 1, 2$, also gets the index $w_j$ on its dedicated private link. Then, Legitimate Receiver $j$, $j = 1, 2$, sets its reconstruction of source $s_j^n$ as $\hat{s}_j^n(w, w_j)$.

*Equivocation Analysis:* The analysis of the equivocation level that is achieved by this coding scheme follows straightforwardly from that in Appendix A for the Helper problem with equivocation constraint of Figure 2a by simply substituting $(S_0^n, S_1^n)$ therein with $(S_1^n, S_2^n)$; and, hence, is omitted for brevity.

## Appendix G
## Proof of Theorem 5

### A. Proof of Converse:

Let $(R, R_1, R_2, D_1, D_2)$ be an achievable rate-distortion tuple for the lossy Gray-Wyner model of Figure 3b. Let $f$, $f_1$ and $f_2$ denote the associated encoding functions at the encoder; and $g_1$ and $g_2$ the decoding functions at the legitimate receivers. That is,

$$W = f(S_0^n, S_1^n, S_2^n) \ , \tag{130}$$

$$W_1 = f_1(S_0^n, S_1^n, S_2^n) \ , \tag{131}$$

$$W_2 = f_2(S_0^n, S_1^n, S_2^n) \ , \tag{132}$$

$$\hat{S}_1^n = g_1(W, W_1) \ , \tag{133}$$

$$\hat{S}_2^n = g_2(W, W_2) \ , \tag{134}$$

with

$$\mathbb{E}d_1^{(n)}(S_1^n, \hat{S}_1^n) \leq nD_1 \quad \text{and} \quad \mathbb{E}d_2^{(n)}(S_2^n, \hat{S}_2^n) \leq nD_2. \tag{135}$$

First, using Fano's inequality, the assumption of lossless reconstruction of the source component $S_0^n$ at both legitimate receivers yield that there exists $\epsilon_n$ with $\lim_{n\to\infty} \epsilon_n = 0$ such that

$$H(S_0^n|W, W_1) \leq n\epsilon_n \ , \tag{136a}$$

$$H(S_0^n|W, W_2) \leq n\epsilon_n, \tag{136b}$$

and, by definition, we have

$$H(S_0^n|W) \geq n\Delta \tag{137}$$

Let, for $i = 1, \ldots, n$, the auxiliary random variable $U_i$ defined as

$$U_i \triangleq (W, S_1^{i-1}, S_2^{i-1}, S_0^{i-1}, S_{0,i+1}^n). \tag{138}$$

First, we lower bound the rate of message $W$ as

$$\begin{aligned}
nR &\geq H(W) && (139a)\\
&\geq I(W;S_0^nS_1^nS_2^n) && (139b)\\
&\geq I(W;S_1^nS_2^n|S_0^n) && (139c)\\
&= \sum_{i=1}^{n} I(W;S_{1,i}S_{2,i}|S_0^n,S_1^{i-1},S_2^{i-1}) && (139d)\\
&\overset{(a)}{=} \sum_{i=1}^{n} I(W,S_0^{i-1},S_{0,i+1}^n,S_1^{i-1},S_2^{i-1};S_{1,i}S_{2,i}|S_{0,i}) && (139e)\\
&\overset{(b)}{\geq} \sum_{i=1}^{n} I(U_i;S_{1,i},S_{2,i}|S_{0,i})\ , && (139f)
\end{aligned}$$

where $(a)$ holds since the sources are memoryless; and $(b)$ follows by substituting using (138).
Next, we lower bound the rate of message $W_j$, $j=1,2$, as

$$\begin{aligned}
nR_j &\geq H(W_j) && (140a)\\
&\geq H(W_j|W) && (140b)\\
&\geq I(W_j;S_0^nS_1^nS_2^n|W) && (140c)\\
&= I(W_j;S_0^n|W)+I(W_j;S_1^nS_2^n|WS_0^n) && (140d)\\
&\overset{(a)}{\geq} H(S_0^n|W)+I(W_j;S_1^nS_2^n|WS_0^n)-n\epsilon_n && (140e)\\
&\overset{(b)}{\geq} n\Delta+I(W_j;S_1^nS_2^n|WS_0^n)-n\epsilon_n && (140f)\\
&\overset{(c)}{\geq} n\Delta+I(\hat{S}_j^n;S_1^nS_2^n|WS_0^n)-n\epsilon_n && (140g)\\
&= n\Delta+\sum_{i=1}^{n} I(\hat{S}_j^n;S_{1,i}S_{2,i}|WS_0^nS_1^{i-1}S_2^{i-1})-n\epsilon_n && (140h)\\
&\overset{(d)}{=} n\Delta+\sum_{i=1}^{n} I(\hat{S}_j^n;S_{1,i}S_{2,i}|S_{0,i},U_i)-n\epsilon_n && (140i)\\
&\geq n\Delta+\sum_{i=1}^{n} I(\hat{S}_{j,i};S_{1,i}S_{2,i}|S_{0,i}U_i)-n\epsilon_n, && (140j)
\end{aligned}$$

where $(a)$ follows by using (142); $(b)$ follows by using (137); $(c)$ holds since $\hat{S}_j^n$ is a deterministic function of the messages $(W,W_j)$; and $(d)$ follows by substituting using (138).
Similarly, for $j\in[1:2]$, we write that

$$\begin{aligned}
n(R+R_j) &\geq H(W)+H(W_1) && (141a)\\
&\geq H(WW_j) && (141b)\\
&\geq I(WW_j;S_0^nS_1^nS_2^n) && (141c)\\
&= I(WW_j;S_0^n)+I(WW_j;S_1^nS_2^n|S_0^n) && (141d)\\
&\overset{(a)}{\geq} I(WW_j;S_0^n)+I(W\hat{S}_j^n;S_1^nS_2^n|S_0^n) && (141e)\\
&\geq H(S_0^n)+I(W\hat{S}_j^n;S_1^nS_2^n|S_0^n)-n\epsilon_n && (141f)
\end{aligned}$$

$$\geq nH(S_0) + \sum_{i=1}^{n} I(W\hat{S}_{j,i}; S_{1,i}S_{2,i}|S_{0,i}S_1^{i-1}, S_2^{i-1}S_0^{i-1}S_{0,i+1}^n) - n\epsilon_n \tag{141g}$$

$$\overset{(a)}{=} nH(S_0) + \sum_{i=1}^{n} I(WS_1^{i-1}S_2^{i-1}S_0^{i-1}S_{0,i+1}^n\hat{S}_{j,i}; S_{1,i}S_{2,i}|S_{0,i}) - n\epsilon_n \tag{141h}$$

$$\overset{(b)}{\geq} nH(S_0) + \sum_{i=1}^{n} I(U_i\hat{S}_{j,i}; S_{1,i}, S_{2,i}|S_{0,i}) \; - n\epsilon_n, \tag{141i}$$

where $(a)$ holds since $\hat{S}_j^n$ is a function of $(W, W_j)$, and $(b)$ follows from that the sources are memoryless.

Finally, a trivial bound on the equivocation can be written as follows:

$$n\Delta \leq H(S_0^n|W) \tag{142a}$$

$$\leq H(S_0^n) \tag{142b}$$

$$= nH(S_0) \; . \tag{142c}$$

The rest of the proof of converse follows using standard single-letterization techniques; and is omitted for brevity.

*B. Proof of Achievability:*

The achievability proof can be outlined as follows. The first stage of the coding scheme consists in compressing the common source $S_0^n$ to be secured through the private links and the common link as well, yielding thus a given leakage at the eavesdropper on this public link. Once $S_0^n$ is decoded at both terminals, it serves as side information available at all terminals to apply a Gray-Wyner code conditioned on it. Such a code consists in compressing jointly the two sources $(S_1^n, S_2^n)$ on the common link with a common description $U^n$, and then, transmitting individual descriptions $\hat{S}_1^n$ and $\hat{S}_2^n$, each on a private link, to refine the common description $U^n$. Since this encoding is performed conditioned on $S_0^n$, a similar argument to the one used in the proof of Theorem 2 is called here, to show that no information is leaked in this second stage of the coding scheme. The codebook generation, encoding and decoding are detailed in the following.

*Codebook Generation:*

i) Assign to each sequence $s_0^n \in \mathcal{T}_{[S_0]}^{(n)}$ independently and at random a pair of indices $(w_0, \bar{w}_0) \in [1 : 2^{nR_0}] \times [1 : 2^{n\bar{R}_0}]$, and an error pair of indices $(w_0, \bar{w}_0) = (1 + 2^{nR_0}, 1 + 2^{n\bar{R}_0})$ to every $s_0^n \notin \mathcal{T}_{[S_0]}^{(n)}$.

ii) For each $s_0^n$, generate $2^{nR}$ sequences $u^n(w)$ where $w \in [1 : 2^{nR}]$, with i.i.d components drawn following $\prod_{i=1}^{n} P_{U|S_0}(u_i(w)|s_{0,i})$.

iii) For each $s_0^n$ and $u^n(w)$, generate $2^{n(R_j-R_0)}$ sequences $\hat{s}_j^n(w, w_j)$ where $w_j \in [1 : 2^{n(R_j-R_0)}]$ following $\prod_{i=1}^{n} P_{\hat{S}_j|US_0}(\hat{s}_{j,i}(w, w_j)|s_{0,i}, u_i(w))$.

*Encoding:* Upon observing $S_0^n$, the encoder:

- Finds the bin indices $(w_0, \bar{w}_0)$ of $s_0^n$.
- Finds a sequence $u^n(w)$ jointly typical with the sources $(s_0^n, s_1^n, s_2^n)$, i.e.

$$\left(u^n(w), s_0^n, s_1^n, s_2^n\right) \in \mathcal{T}_{[US_0S_1S_2]}^{(n)} \; . \tag{143}$$

- Then, for $j \in \{1,2\}$, finds a sequence $\hat{s}_j^n(w, w_j)$ such that:

$$\left(u^n(w), \hat{s}_j^n(w, w_j), s_0^n, s_1^n, s_2^n\right) \in \mathcal{T}^{(n)}_{[U\hat{S}_j S_0 S_1 S_2]} . \tag{144}$$

- Transmits the indices $w_0$ and $w_j$ on each of the private links, then transmit $w$ and $w_{\bar{0}}$ on the common link.

*Estimation:* Each of the decoders first recovers the sequence $s_0^n$ and the sequence $u^n(w)$. Upon decoding both sequences, they each set their reconstruction as $\hat{s}_j^n(w, w_j)$.

The encoding and decoding are successful provided that $n$ is large and the following set of constraints is verified:

$$R_0 + \bar{R}_0 \geq H(S_0) \tag{145a}$$

$$R - \bar{R}_0 \geq I(U; S_1 S_2 | S_0) \tag{145b}$$

$$R_1 - R_0 \geq I(\hat{S}_1; S_1 S_2 | U S_0) \tag{145c}$$

$$R_2 - R_0 \geq I(\hat{S}_1; S_1 S_2 | U S_0) . \tag{145d}$$

*Equivocation Analysis:* The analysis of the equivocation level that is achieved by this coding scheme follows straightforwardly from that in Appendix A for the Helper problem with equivocation constraint of Figure 2b by simply substituting $S_1^n$ therein with $(S_1^n, S_2^n)$; and, hence, is omitted for brevity. The resulting bound on the equivocation level at the eavesdropper about the source $s_0^n$ is given by:

$$\Delta \leq \min\{R_0, H(S_0)\} . \tag{146}$$

*Fourier-Motzkin Elimination:* Summarizing, the tuple $(R, R_1, R_2, D_1, D_2, \Delta)$ is achievable if there exists $(R_0, \bar{R}_1)$ such that

$$R_0 + \bar{R}_0 \geq H(S_0) \tag{147a}$$

$$R - \bar{R}_0 \geq I(U; S_1 S_2 | S_0) \tag{147b}$$

$$R_1 - R_0 \geq I(\hat{S}_1; S_1 S_2 | U S_0) \tag{147c}$$

$$R_2 - R_0 \geq I(\hat{S}_1; S_1 S_2 | U S_0) \tag{147d}$$

$$\Delta \leq \min\{H(S_0), R_0\} \tag{147e}$$

$$0 \leq R_0 \leq R_1 \tag{147f}$$

$$0 \leq \bar{R}_0 \leq R \tag{147g}$$

Using Fourier-Motzkin elimination to successively project out $R_0$, and $\bar{R}_0$, we complete the proof of Theorem 5.